\documentclass[aps,prc,floatfix,preprint]{revtex4-1}
\usepackage{graphicx}
\usepackage{epsfig}
\usepackage{color}
\usepackage{rotating}
\usepackage[normalem]{ulem}  

\begin{document}
\title{Transport properties of nuclear pasta phase with quantum molecular dynamics}
\author{Rana Nandi}
\email[]{rana.nandi@tifr.res.in}
\affiliation{Tata Institute of Fundamental Research, Mumbai-400005, India}
\author{Stefan Schramm}
\email[]{schramm@fias.uni-frankfurt.de}
\affiliation{Frankfurt Institute for Advanced Studies, 60438 Frankfurt am Main, Germany}
 
\begin{abstract}
We study the transport properties of nuclear pasta for a wide range of density, temperature and proton fractions, relevant
for different astrophysical scenarios adopting a quantum molecular dynamics model. In particular, we estimate the values of shear
viscosity as well as electrical and thermal conductivities by calculating the static structure factor $S(q)$ using simulation data.
In the density and temperature range where the pasta phase appears, the static structure factor shows irregular behavior. 
The presence of a slab phase greatly enhances the peak in $S(q)$. However, the effect of irregularities in $S(q)$ on the transport 
coefficients is not very dramatic. The values of all three transport coefficients are found to have the same orders of magnitude as 
found in theoretical calculations for the inner crust matter of neutron stars without the pasta phase and therefore, is in contrast to 
earlier speculations that a pasta layer might be highly resistive, both thermally and electrically.
\end{abstract}

\maketitle  
 
\section{Introduction}\label{sec:intro}
Nuclei present in the neutron star crust at densities well below the saturation density are spherical in shape and form crystalline structures 
to minimize the Coulomb energy. When nuclei are about to dissolve into uniform matter at higher densities, various exotic structures such as 
cylindrical and slab-shaped nuclei, cylindrical and spherical bubbles, etc., collectively known as nuclear ``pasta'' may appear \citep{Ravenhall83, Hashimoto84}.

The pasta phase plays a very important role in understanding various astrophysical phenomena. For example, the neutrino transport in core-collapse supernovae is 
greatly affected by the pasta phase \citep{Horowitz04a, Horowitz04b}. In the crustal matter of neutron stars electrons are the main carriers of charge and momentum.
Therefore, electron-pasta scattering is supposed to affect significantly the transport properties like shear viscosity, thermal and electrical conductivities
of the crust. The thermal conductivity of the crust is the key factor to understand the cooling behavior of neutron stars \citep{Newton13, Horowitz15}
and the reduced electrical conductivity of the pasta could be crucial to explain the decay of magnetic field in them \citep{Pons13}. On the other hand, the shear
viscosity of the crust is essential for the calculation of viscous damping of $p$-modes of neutrons stars \citep{Chugonov05}. The pasta phase might also play a
crucial role to understand the mechanism of pulsar glitches \citep{Lorenz93, Watanabe07}.

The pasta phase has been studied extensively by static methods such as liquid-drop models \citep{Lorenz93, Watanabe00, Watanabe01, Watanabe03a}, Thomas-Fermi approximations
\citep{Oyamatsu93, Lassaut87}, and the Hartee-Fock method \citep{Gogelein07, Newton09} as well as using dynamical approaches like classical 
\citep{Horowitz04a,Horowitz04b,Horowitz08,Horowitz09,Chugonov10,Horowitz15,Schneider13,Schneider14,Dorso12,Molinelli14} and quantum 
molecular dynamics simulations \citep{Maruyama98, Watanabe03b, Watanabe04, Watanabe05, Watanabe09, Nandi16, Nandi17, Schramm17a, Schramm17b} and the
time-dependent Hartree-Fock method \citep{Schuetrumpf13,Fattoyev17}. Although several authors \citep{Flowers76, Nandkumar83, Potekhin99, Chugonov05} have studied transport 
properties of the crust, very few studies employed dynamical approach. At first, \citet{Horowitz04a, Horowitz04b} and recently 
\citet{Alcain14} studied neutrino-pasta scattering adopting classical molecular dynamics.  \citet{Horowitz08, Horowitz15} also estimated the values
of the shear viscosity and thermal conductivity of the pasta phase but only for a few densities and a single temperature and proton fraction. In this work we calculate
shear viscosity, electrical and thermal conductivities for a wide range of density ($0.1-0.6\rho_0$) and temperature ($T=0-5$ MeV)
and three values of the proton fraction ($Y_p$) relevant for different astrophysical scenarios, using a model of quantum molecular dynamics (QMD).

The article is organized as following. The QMD model and the method to calculate the transport coefficients are given in Sec. \ref{sec:Formalism}.
In Sec. \ref{sec:simulation}, we describe the simulation procedure. We present the results with discussion in Sec. \ref{sec:results}.
Finally, we summarize and conclude in Sec. \ref{sec:summary}.

\section{Formalism}
\label{sec:Formalism}
\subsection{QMD}
In the QMD approach the total ${\cal N}$-nucleon wave function $\Psi(\{\bf r\})$ is assumed to be a direct product of single-nucleon wave functions \citep{Maruyama98}:
\begin{equation}
 \Psi(\{{\bf r}\}) =  \prod_i^{\cal N} \psi_i({\bf r}) ,
\end{equation}
where the single-nucleon wave functions $\psi_i(\bf r)$ are represented by Gaussian wave packets (we set $\hbar=c=k_B=1$):
\begin{equation}
 \psi_i({\bf r}) = \frac{1}{(2\pi C_W)^{3/4}} \exp\left[-\frac{({\bf r - R_i})^2}{4C_W}+ i \,{\bf r\cdot P_i}\right]-
\end{equation}
$C_W$ is the width of the Gaussian wave packets and the centers of the position and momentum of the wave packet $i$ are denoted by ${\bf R}_i$ and ${\bf P}_i$, 
respectively.

Here we adopt the QMD Hamiltonian developed by \citet{Chikazumi01}, for the simulation of
nuclear matter at sub-saturation densities. The Hamiltonian has several components:
\begin{equation}
 {\cal H}=K+V_{\rm Pauli}+V_{\rm Skyrme}+V_{\rm sym}+V_{\rm MD}+V_{\rm Surface}+V_{\rm Coul},
\end{equation}
where $K$ is the kinetic energy, $V_{\rm Pauli}$ is the  Pauli potential introduced \citep{Dorso87} to include the Pauli exclusion
principle phenomenologically, $V_{\rm Skyrme}$ is the nucleon-nucleon potential similar to Skyrme-like interactions, $V_{\rm sym}$ is 
the isospin-dependent potential related to the symmetry energy, $V_{\rm MD}$ is the momentum-dependent potential incorporated as
Fock terms of Yukawa-type interactions, $V_{\rm Surface}$ is the potential that depends on the density gradient
and finally, $V_{\rm Coul}$ is the Coulomb potential. 
The explicit expressions for all the terms are given below \citep{Chikazumi01}:
\begin{widetext}
\begin{eqnarray}
  K  &=& \sum_i \frac{\bf P_{\it i}^{2}}{2 m_{i}}\ ,\label{kin}\\  
  V_{\rm Pauli} &=& 
  \frac{C_{\rm P}}{2}\
  \left( \frac{1}{q_0 p_0}\right)^3
  \sum_{i, j(\neq i)} 
  \exp{ \left [ -\frac{({\bf R}_i-{\bf R}_j)^2}{2q_0^2} 
          -\frac{({\bf P}_i-{\bf P}_j)^2}{2p_0^2} \right ] }\
  \delta_{\tau_i \tau_j} \delta_{\sigma_i \sigma_j}\ ,\label{pauli}\\
  V_{\rm Skyrme} &=&
  {\alpha\over 2\rho_0}\sum_{i, j (\neq i)}
  \rho_{ij}
  +  {\beta\over (1+\tau)\ \rho_0^{\tau}}
  \sum_i \left[ \sum_{j (\neq i)} 
                     \tilde{\rho}_{ij}  \right]^{\tau}\ ,
                   \label{skyrme}\\
   V_{\rm sym} &=&
  {C_{\rm s}\over 2\rho_0} \sum_{i , j(\neq i)} \,
  ( 1 - 2 | \tau_i - \tau_j | ) \ \rho_{ij} \\
  V_{\rm MD}  &=&  {C_{\rm ex}^{(1)} \over 2\rho_0} \sum_{i , j(\neq i)} 
      {1 \over 1+\left[{{\bf P}_i-{\bf P}_j \over \mu_1}\right]^2} 
      \ \rho_{ij}
     +   {C_{\rm ex}^{(2)} \over 2\rho_0} \sum_{i , j(\neq i)} 
      {1 \over 1+\left[{{\bf P}_i-{\bf P}_j \over  \mu_2}\right]^2} 
      \ \rho_{ij}\ ,\label{md}\\
  V_{\rm Surface} &=&  \frac{V_{\rm SF}}{2\rho_0^{5/3}} \sum_{j (\neq i)}\int d{\bf r}\nabla\rho_i({\bf r})\cdot \nabla\rho_j({\bf r}) \, \\ 
  V_{\rm Coul} &=&
  {e^2 \over 2}\sum_{i , j(\neq i)}
  \left(\tau_{i}+\frac{1}{2}\right) \, \left(\tau_{j}+\frac{1}{2}\right)
  \int\!\!\!\!\int d^3{\bf r}\,d^3{\bf r}^{\prime} 
  { 1 \over|{\bf r}-{\bf r}^{\prime}|} \,
  \rho_i({\bf r})\rho_j({\bf r}^{\prime})\ ,\label{coulomb}
\end{eqnarray}
\end{widetext}
where $\rho_0$($=0.165{\rm fm}^{-3}$) denotes the normal nuclear matter density, $\sigma_{i}$ and $\tau_{i}$ ($1/2$ for protons and $-1/2$ for neutrons) are the
nucleon spin and isospin, respectively and the overlap between single-nucleon densities are represented by $\rho_{ij}$ and $\tilde{\rho}_{ij}$:
\begin{equation}
  \rho_{ij} \equiv \int { d^3{\bf r} \rho_i({\bf r})
  \rho_j({\bf r}) }\ ,\quad \tilde{\rho}_{ij} \equiv \int { d^3{\bf r} \tilde{\rho_i}({\bf r})\tilde{ \rho_j}({\bf  r})}\ ,
  \label{rhoij}
\end{equation}
with the single-nucleon densities:
\begin{eqnarray}
  \rho_i({\bf r}) & = & \left| \psi_{i}({\bf r}) \right|^{2}
  = \frac{1}{(2\pi C_W)^{3/2}}\ \exp{\left[
                - \frac{({\bf r} - {\bf R}_i)^2}{2C_W} \right]}\ ,\quad \\
  \tilde{\rho_i}({\bf r}) & = &
  \frac{1}{(2\pi \tilde{C}_W)^{3/2}}\ \exp{\left[
                - \frac{({\bf r} - {\bf R}_i)^2}{2\tilde{C}_W} \right]}\ , 
 \end{eqnarray}
where
\begin{equation}
  \tilde{C}_W = \frac{1}{2}(1+\tau)^{1/ \tau}\ C_W 
\end{equation}
is the modified width of the Gaussian wave packet introduced to adjust the effect of density-dependent terms 
\citep{Maruyama98}. The values of the parameters of the Hamiltonian are shown in Table \ref{tab:parameter}. They are determined to reproduce
the saturation properties of nuclear matter as well as ground state properties of finite nuclei.
\begin{table}
\begin{center}
\caption{Parameter set for the interaction \citep{Maruyama98}}
{\small \begin{tabular}{cccc}
\hline\hline 
& $C_{\rm P}$ (MeV) &\qquad\qquad 115.0 &\\
& $p_{0}$ (MeV/$c$) &\qquad\qquad 120.0 &\\
& $q_{0}$ (fm) &\qquad\qquad 2.5 &\\
& $\alpha$ (MeV) &\qquad\qquad $-121.9$&\\
& $\beta$ (MeV) &\qquad\qquad 197.3 &\\
& $\tau$ &\qquad\qquad 1.33333 &\\
& $C_{\rm s}$ (MeV) &\qquad\qquad 25.0 &\\
& $C_{\rm ex}^{(1)}$ (MeV) &\qquad\qquad $-258.54$ &\\
& $C_{\rm ex}^{(2)}$ (MeV) &\qquad\qquad 375.6 &\\
& $\mu_1$ (fm$^{-1}$) &\qquad\qquad 2.35 &\\
& $\mu_2$ (fm$^{-1}$) &\qquad\qquad 0.4 &\\
& $V_{\rm SF}$ (MeV) & \qquad\qquad 20.68 &\\
& $C_W$ (fm$^2$) &\qquad\qquad 1.95 &\\
\hline\hline 
\end{tabular}}
\label{tab:parameter}
\end{center}
\end{table}

In order to obtain the equilibrium configuration we use the QMD equations of motion with friction terms \citep{Maruyama98}:
\begin{eqnarray}
 {\bf\dot{R}_i } &=& \frac{\partial H}{\partial {\bf P_i}} - \mu_R\frac{\partial H}{\partial {\bf R_i}},\nonumber\\ 
 {\bf\dot{P}_i } &=& -\frac{\partial H}{\partial {\bf R_i}} - \mu_P\frac{\partial H}{\partial {\bf P_i}},\label{eom}
\end{eqnarray}
where the damping coefficients $\mu_R$ and $\mu_P$ are positive definite and are related to the relaxation time scale.

The QMD Hamiltonian adopted here contains momentum-dependent interactions ($V_{\rm Pauli}$  and $V_{\rm MD}$). Therefore, we cannot use the
usual expressions for the instantaneous temperature:
\begin{equation}
\frac{3}{2}\, T=\frac{1}{\cal N}\sum_{i=1}^{\cal N} \frac{{\bf P}_i^2}{2m_i}.
\label{eq:Tkin}
\end{equation}
Instead, we use the following definition of an effective temperature \citep{Chikazumi01,Watanabe04}:
\begin{equation}
 \frac{3}{2}\, T_{\rm eff} = \frac{1}{\cal N}\sum_{i=1}^{\cal N}\frac{1}{2}{\bf P}_i\cdot \frac{\partial{\cal H}}{\partial{\bf P}_i},
 \label{eq:Teff}
\end{equation}
This reduces to the usual definition of Eq. (\ref{eq:Tkin}) when the Hamiltonian does not contain any momentum-dependent interaction.
It was shown \citep{Watanabe04} by performing Metropolis Monte Carlo simulations that $T_{\rm eff}$ is consistent with the 
temperature in the Boltzmann statistics.

In order to perform simulations at a specified temperature ($T_{\rm set}$) we adopt the Nos\'{e}-Hoover thermostat \citep{Nose,Hoover,Allen} following the 
prescription of  \citet{Watanabe04}. The extended Hamiltonian that includes the thermostat is given by:
\begin{equation}
  {\cal H}_{\rm Nose} = \sum_{i=1}^{\cal N}\frac{{\bf P}_i^2}{2m_i} + {\cal U}(\{{\bf R}_i\},\{{\bf P}_i)\} + \frac{s^2p_s^2}{2}\,
  + g\frac{{\rm ln}\,s}{\beta}
\end{equation}
where the potential ${\cal U} (\{{\bf R}_i\}),\{{\bf P}_i\})  = {\cal H} - K$ depends on both positions and momenta, 
$s$ denotes the extended variable for the thermostat, $p_s$ is the conjugate momentum corresponding to $s$, $Q$ represents the effective ``mass'' associated 
with $s$ and takes a value $\sim10^8\, {\rm MeV\,fm}^2$, $g=3{\cal N}$ is necessary to generate the canonical ensemble, and $\beta=1/T_{\rm set}$.
The extended system evolves according to the following equations of motion:
\begin{eqnarray}
 {\bf\dot{R}_i} &=& \frac{{\bf P}_i}{m_i} + \frac{\partial{\cal U}}{\partial {\bf P}_i}\,\\
 {\bf\dot{P}_i} &=& -\frac{\partial {\cal U}}{\partial {\bf R_i}} - \xi{\bf P}_i,\\
 \dot{\xi} &=& \frac{1}{Q}\left[\sum_{i=1}^{\cal N}\left(\frac{{\bf P}_i}{m_i}+{\bf P}_i\cdot\frac{\partial{\cal U}}{\partial{\bf P}_i} \right)
 -\frac{g}{\beta}\right]\,\\
 \dot{s}/s &=& \xi \, 
\end{eqnarray}
where  $\xi(=sp_s/Q$) acts as thermodynamic friction coefficient. During the simulation
 $T_{\rm eff}$ fluctuates around $T_{\rm set}$, whereas ${\cal H}_{\rm Nose}$ should remain conserved.

\subsection{Transport properties}
In the astrophysical conditions considered here electrons are the most important carriers of charge and momentum and therefore the transport properties 
like shear viscosity ($\eta$), thermal and electrical conductivities ($\sigma$, $\kappa$) are determined from scattering of electrons off
ions \citep{Chugonov05,Nandkumar83}. Following \citet{Horowitz08} we choose to work in nucleon coordinates instead of ion coordinates, 
as the former are more suitable for complicated pasta phases where the identification of ions is not always possible.
Then the transport coefficients for degenerate electrons can be written as
\citep{Potekhin99, Chugonov05, Nandkumar83}:
\begin{eqnarray}
\kappa &=& \frac{\pi k_F^3T}{12e^4 m_e^{*2}\Lambda_{ep}^{\kappa}}, \label{eq:tc}\\
\sigma &=& \frac{k_F^3}{4\pi e^2m_e^{*2}\Lambda_{ep}^{\sigma}},\label{eq:ec}\\
\eta &=& \frac{k_F^5}{60\pi e^4m_e^{*2}\Lambda_{ep}^\eta},\label{eq:sv}
\end{eqnarray}
where $k_F$ is the electron Fermi momentum, $m_e^*=\sqrt{k_F^2 + m_e^2}$ is the electron effective mass, $m_e$ is the electron rest mass and
$\Lambda_{ep}$s are the Coulomb logarithms that describe electron-proton scattering:
\begin{eqnarray}
 \Lambda_{ep}^\kappa &=& \int_0^{2k_F}\frac{dq}{q}\frac{F_N(q)^2}{\varepsilon(q)^2}S_p(q)\left(1-\frac{q^2}{4m_e^{*2}}\right), \label{eq:clc}\\
 \Lambda_{ep}^\sigma &=& \Lambda_{ep}^\kappa, \label{eq:clc1}\\
 \Lambda_{ep}^\eta &=& \int_0^{2k_F}\frac{dq}{q}\frac{F_N(q)^2}{\varepsilon(q)^2}S_p(q)\left(1-\frac{q^2}{4k_F^2}\right)
                      \left(1-\frac{q^2}{4m_e^{*2}}\right), \label{eq:clv}
\end{eqnarray}
where $q$ is the momentum transfer, $F_N(q)=e^{-q^2C_W/2}$ is the form factor of the Gaussian wave-packet of nucleons,
$\varepsilon(q)$ ($\approx 1,$ if $q$ is not very small) is the static longitudinal dielectric function of the electron 
gas \citep{Jancovici62}, and $S_p(q)$ is the static structure factor that describes correlations between protons. 
The static structure factor is calculated from the autocorrelation function
\begin{equation}
 S_p(q) = \frac{1}{N_p}<\rho_p(q,t)^*\rho_p(q,t)> \label{eq:sf}
\end{equation}
of the Fourier transform of the proton number density
\begin{equation}
 \rho_p(q,t) = \sum_{j=1}^{N_p}e^{i{\bf q}\cdot {\bf R}_j(t)} \label{eq:nd}.
\end{equation}
Here $N_p$ is the number of protons in the system, ${\bf R}_j(t)$ is the center of position of the $j$-th wave packet at time $t$. The average in
Eq. (\ref{eq:sf}) is taken over time $t$ in the simulation and the directions of ${\bf q}$.
Due to the periodic boundary conditions employed in the MD simulation ${\bf q}$ takes discrete values:
\begin{equation}
 {\bf  q} = \frac{2\pi}{L}(l,m,n), \label{eq:qdis}
\end{equation}
where $l\,,m$ and $n$ are integers and $L$ is the length of the cubic simulation box.

\section{Simulation}\label{sec:simulation}
Adopting the theoretical framework outlined in the previous section we perform QMD simulations of nuclear matter
for a wide range of density ($\rho=0.1-0.6\rho_0$) and temperature ($T=0-5$ MeV). We investigate symmetric nuclear matter
(proton fraction $Y_p=0.5$) important for heavy-ion collisions
as well as asymmetric nuclear matter with $Y_p=0.3$, typical for a supernova environment and $Y_p=0.1$, relevant for 
neutron stars. We take $\cal N$ nucleons in a cubic box the size of which is determined from $\cal N$ and the chosen density $\rho$ as 
$L=({\cal N}/\rho)^{1/3}$.
Periodic boundary conditions are imposed to simulate infinite matter. 
The number of protons (neutrons) with spin-up is taken to be equal to that of protons (neutrons) with spin-down. To calculate
the Coulomb interaction we employ the Ewald method \citep{Watanabe03b,Allen}, where electrons are considered to form  a uniform background and
make the system charge neutral. From eq. (\ref{eq:qdis}) we see that the minimum value of $q$ is determined by the box size
$q_{\rm  min}=2\pi/L$. To keep $q_{\rm min}$ the same at all densities we increase $\cal N$ with density as shown in Table \ref{tab:N}.
\begin{table}
 \centering
 \caption{Simualtion data}\label{tab:N}
 \begin{tabular}{ccccc}
 \hline
  $\rho/\rho_0$ & $\cal N$ &        $T=0$          & \multicolumn{2}{c}{$T=1-5$ MeV} \\
                &          & Simulation time (fm/c)& Simulation time(fm/c) & $N_{\rm conf}$ \\               
  \hline\hline
  0.1  &  4096 & $2.00\times10^5$  & $2.50\times 10^5$ & 10140\\
  0.2  &  8192 & $1.60\times10^5$  & $2.00\times 10^5$ &  8040\\
  0.3  & 12288 & $1.60\times10^5$  & $1.75\times 10^5$ &  6940\\
  0.4  & 16384 & $1.80\times10^5$  & $1.50\times 10^5$ &  5640\\
  0.5  & 20480 & $1.80\times10^5$  & $1.00\times 10^5$ &  3340\\
  0.6  & 24576 & $1.50\times10^5$  & $7.00\times 10^4$ &  2040\\ 
  \hline\hline
 \end{tabular}
\end{table}

As an initial configuration we distribute nucleons randomly in phase space. Then with the help of the Nos\'{e}-Hoover 
thermostat we equilibrate the system at $T\sim20$ MeV for about $2000$ fm/c. To achieve the ground state
configuration we then slowly cool down the system in accordance with the damped equations of motion (Eqs. \ref{eom}) until the 
change in the energy per nucleon ($E/{\cal N}$) is less than 1 keV in $10000$ fm/c i.e. 
\begin{equation}
 E/{\cal N}{\big|}_{t+10000\, {\rm fm/c}} - E/{\cal N}{\big|}_t < 1\, {\rm keV}.\label{eq:convg}
\end{equation}
We have listed the simulation time needed to achieve the convergence in the second 
column of Table \ref{tab:N}.
In order to obtain nuclear matter configurations at a finite temperature $T_{\rm set}$
we cool down the system until $T$ reaches $\sim 5$ MeV. Then the system is relaxed for $5000$ fm/c at the desired temperature
$T_{\rm set}$ with the help of the thermostat and, finally, it is further relaxed without the thermostat. All the measurements are
taken at this microcanonical  stage of simulation. Trajectories are stored at every 20 fm/c. In the third column of Table \ref{tab:N} we have shown the 
total simulation time for all the densities considered here. As with increasing density the number of particles increases the simulation becomes
slower. Therefore, we run the simulation for shorter duration with increasing density. Fortunately, this is not a big problem as at higher densities 
the system equilibrates much faster as also found earlier \citep{Horowitz04b}.

\section{Results}\label{sec:results}
In this section we present our results for all densities, proton fractions and temperatures. We first calculate the static
structure factors. Using these results  we then calculate transport coefficients.

\subsection{Static structure factor}
We calculate the static structure factors for protons $S_p(q)$, using Eqn. (\ref{eq:sf}). For $T>0$ the average is taken over all the configurations 
in the final stage of relaxation. The fourth column of Table \ref{tab:N} shows the number of configurations ($N_{\rm conf}$) used for various densities.
To improve the statistics the average is taken also over the directions of $\bf q$. However, for $T=0$ we consider only the final configuration
as the particles do not move at all when $t \gtrsim 10^5$ fm/c.
Similarly, we also calculate static structure factors for
neutrons $S_n(q)$, which are important for the study of neutrino transport in core-collapse supernovae \citep{Horowitz04b}.

\begin{figure}
 \begin{tabular}{cccccc}
  \includegraphics[width=0.32\textwidth]{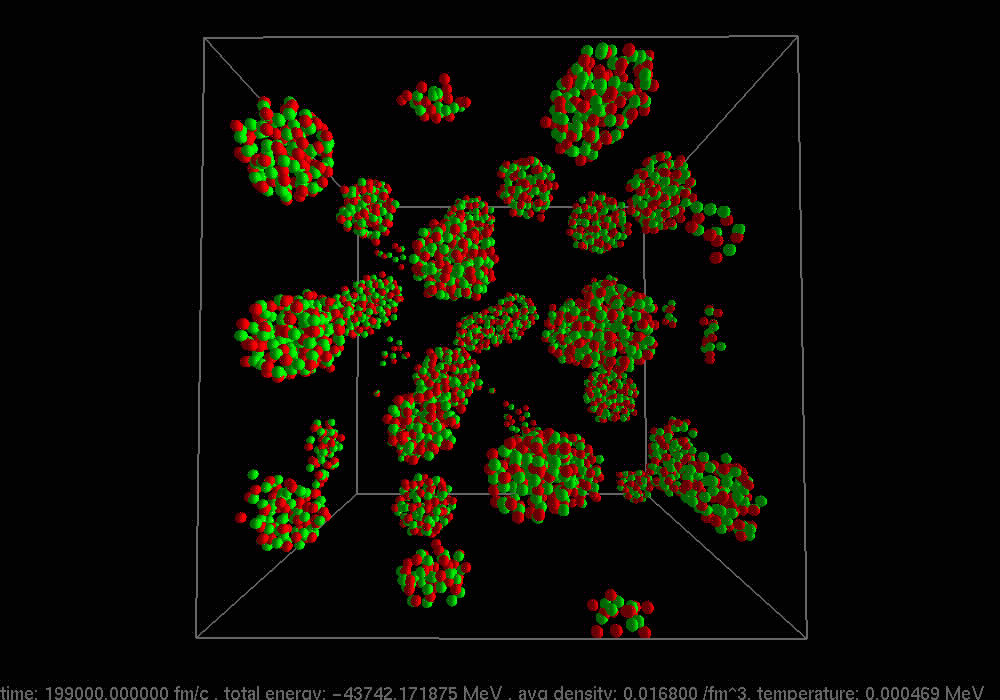}&
  \includegraphics[width=0.32\textwidth]{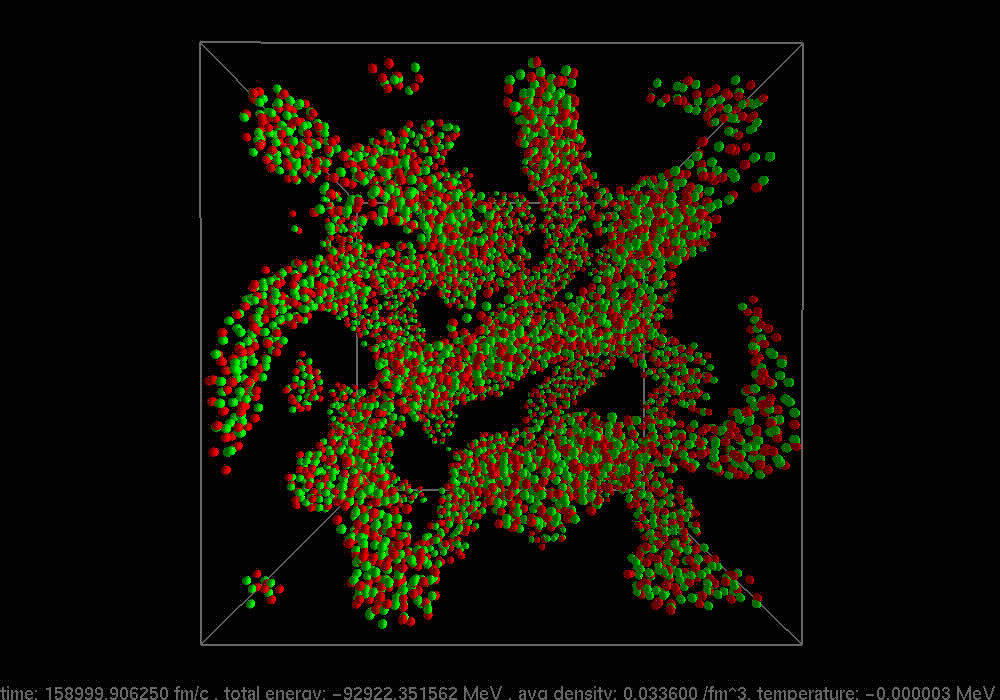}&
  \includegraphics[width=0.32\textwidth]{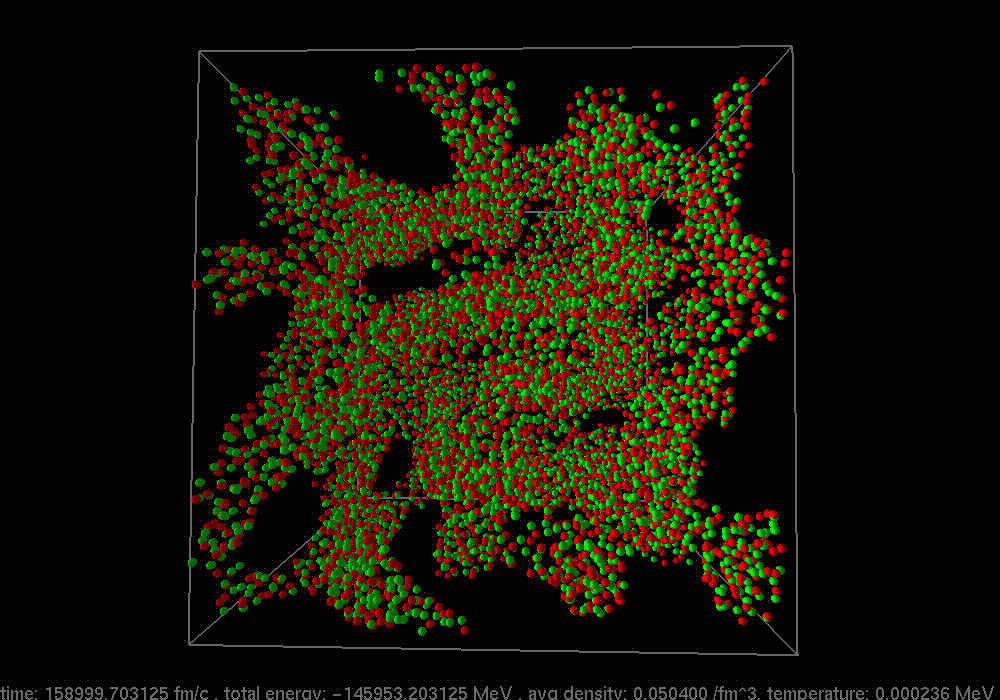}\\
  \includegraphics[width=0.32\textwidth]{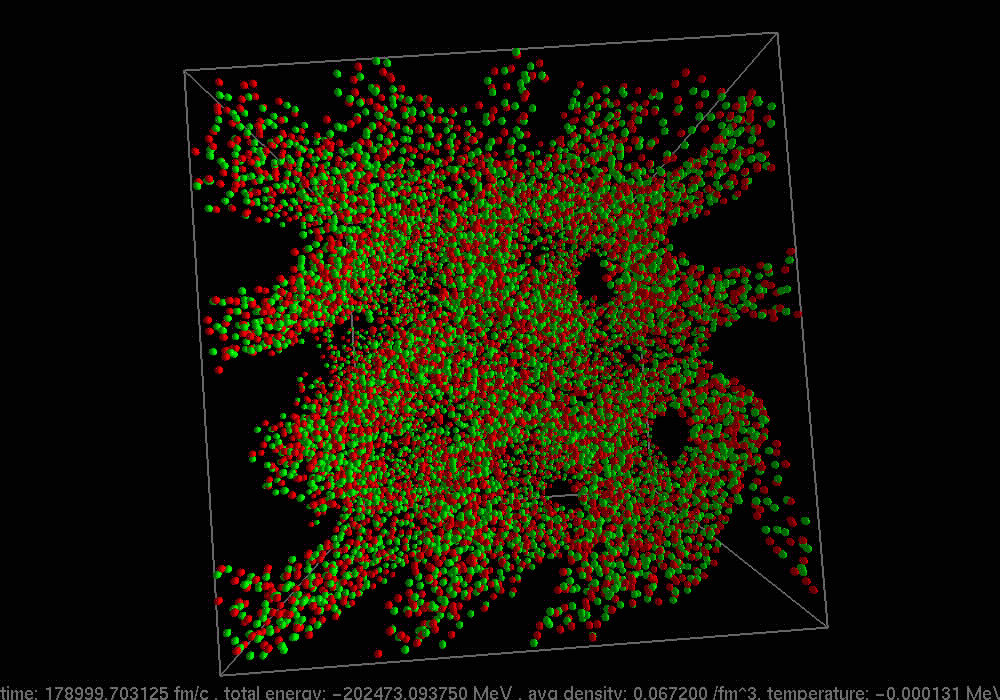}&
  \includegraphics[width=0.32\textwidth]{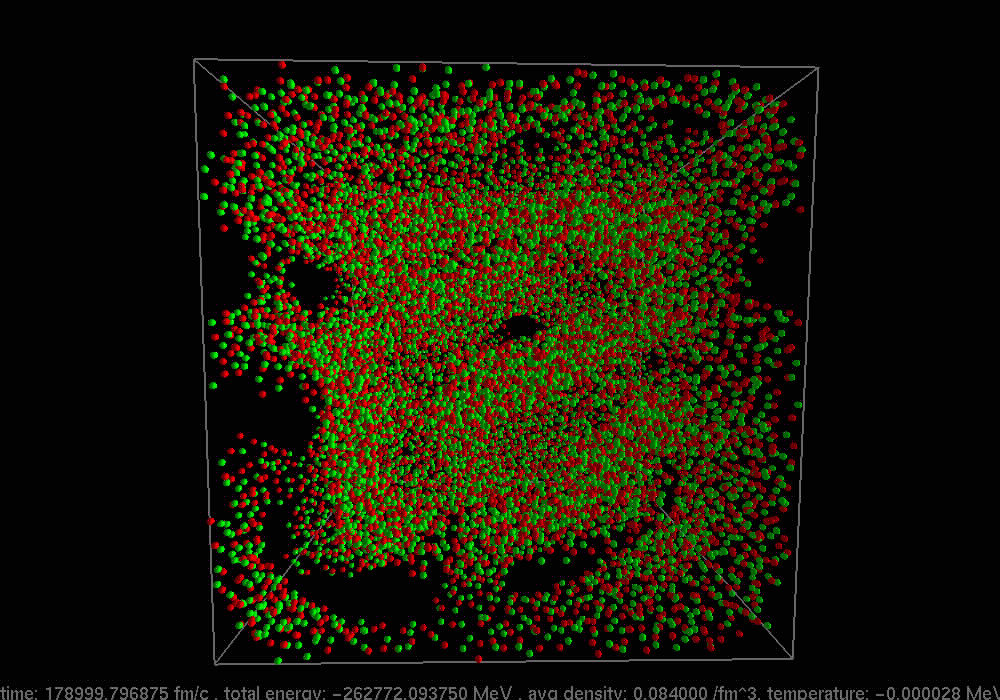}&
  \includegraphics[width=0.32\textwidth]{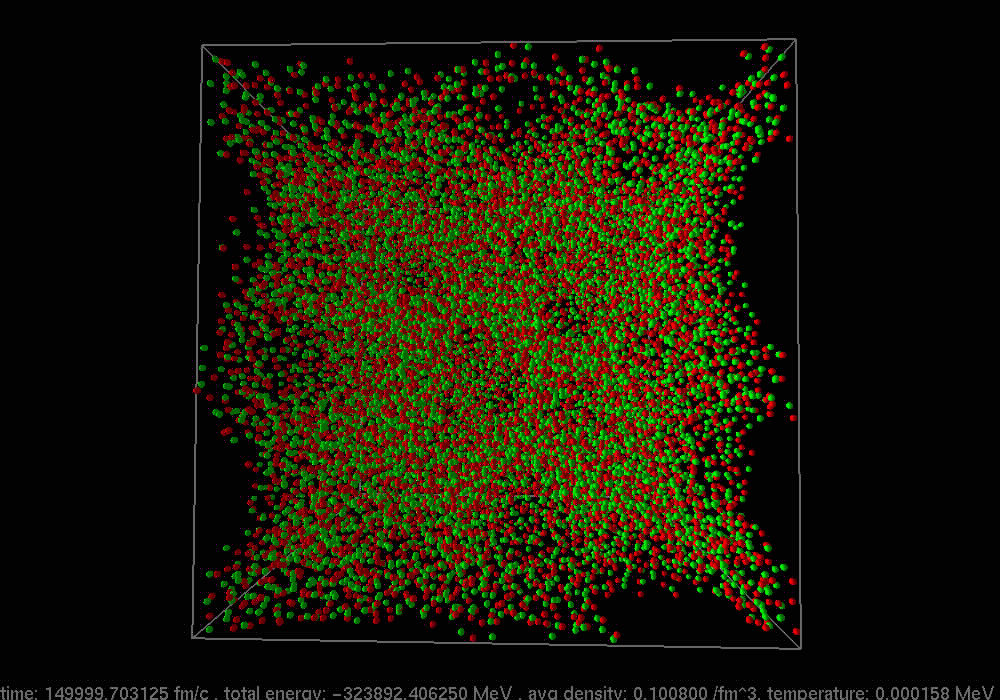}\\
 \end{tabular}
\caption{(Color online) simulation snapshots at $\rho=0.1-0.6\rho_0$, $Y_p=0.5$ and $T=0$. Number of nucleons changes with density according to Table 
\ref{tab:N} keeping the box size fixed to 62.47 fm. Green/gray (red/dark gray) spheres represent neutrons (protons).}
\label{fig:snapshots_xp50_T0}
\end{figure}
Before presenting the results of static structure factors we first show  the simulation snapshots for the density range $\rho=0.1-0.6\rho_0$ and $T=0$ for
symmetric nuclear matter in Fig. \ref{fig:snapshots_xp50_T0}. The snapshot at $\rho=0.1\rho_0$ shows spherical clusters with well-defined surfaces. 
At $\rho=0.2\rho_0$, we are already in pasta phase as elongated spaghetti-like shapes appear. At $\rho=0.3\rho_0$, these bent rods begin to merge and 
at $\rho=0.4\rho_0$ onwards, we obtain complicated bubble shapes.

\begin{figure}
 \begin{center}
 \includegraphics[width=0.7\textwidth,angle=-90]{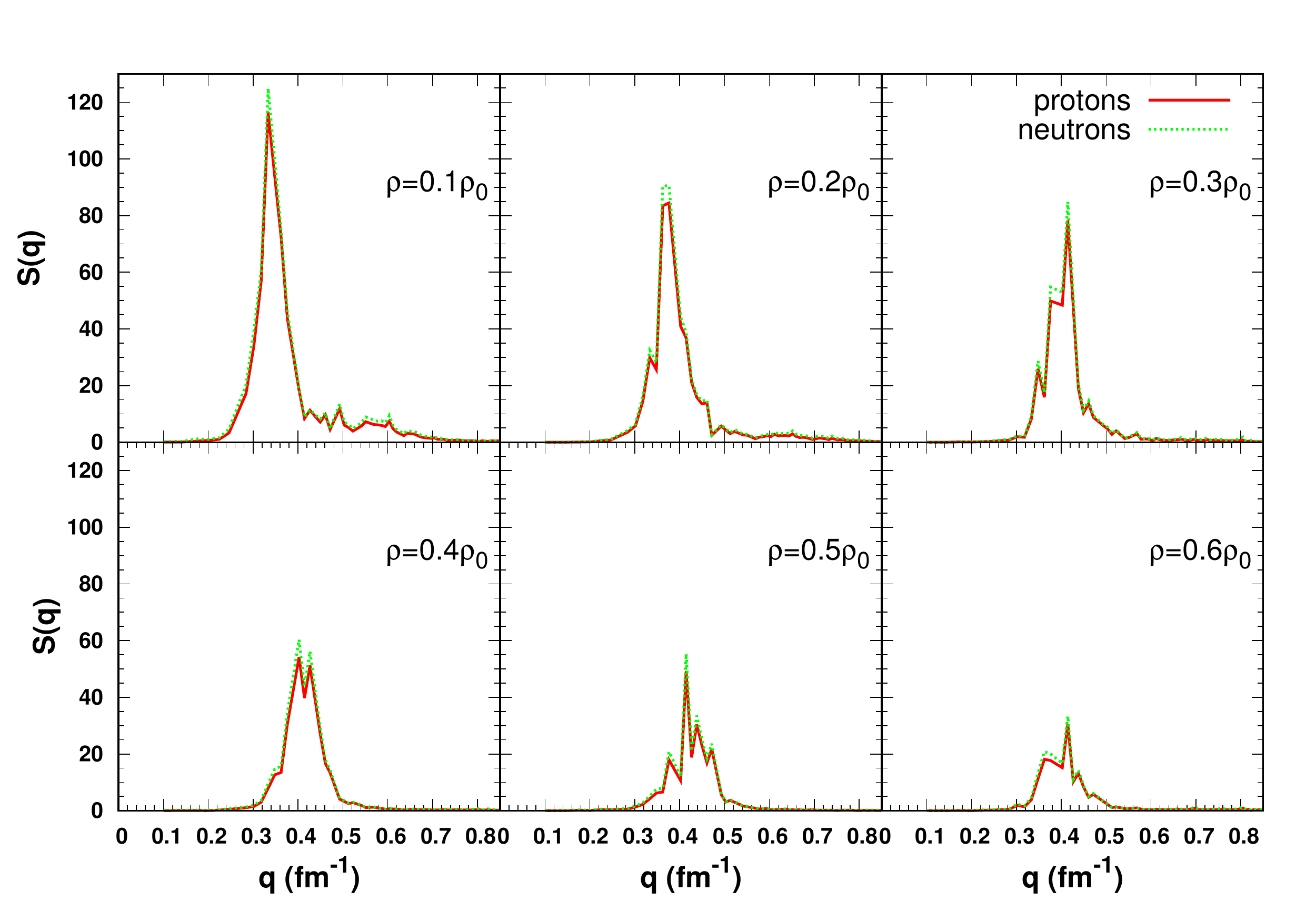}
 \caption{Static structure factor vs momentum transfer ($q$) for protons (solid line) and neutrons (dashed line) at $T=0$, $Y_p=0.5$ and
 $\rho=0.1-0.6\rho_0$. }
 \label{fig:sq_allnb_xp50_T0}
 \end{center}
\end{figure}
In Fig. \ref{fig:sq_allnb_xp50_T0}, we show the static structure factors $S(q)$ for both protons (solid line) and neutrons (dashed line) as a function
of momentum transfer ($q$) for nuclear matter with $Y_p=0.5,$ $T=0$ and $\rho=0.1-0.6\rho_0$.
From the figure it is seen that with increasing density
the height of the peak  $S(q_{\rm peak})$ decreases and the location of the peak ($q_{\rm peak}$) increases till $\rho=0.3\rho_0$ and does not change much thereafter.
The $S(q_{\rm peak})$ is proportional but not equal to the number of particles in the cluster because of the nuclear form factor \citep{Horowitz08}
defined as
\begin{equation}
 F_{p,n}(q) = \frac{1}{N_{p,n}}\int d{\bf r}\,e^{i{\bf q}\cdot{\bf r}}\rho_{p,n}(r),
\end{equation}
where, $F_p(q)$ and $F_n(q)$ denote nuclear form factors containing $N_p$ protons and $N_n$ neutrons and
$\rho_{p,n}(r)$ are the corresponding densities inside a nuclear cluster. The form factor reduces $S(q)$ at high $q$, whereas at low $q$ the reduction
is caused by the screening effects of other ions \citep{Horowitz08}. As the density increases the cluster gets bigger and closer.  Although there 
are more particles in the cluster the $S(q_{\rm peak})$ decreases with density. This happens because the form factor is more effective for larger clusters and the
screening effect is more efficient at higher densities.
The location of the peak is related to the average distance between clusters. As the density increases from $0.1\rho_0$ to $0.3\rho_0$ the nuclear clusters
come closer to give higher values of $q_{\rm peak}$. There is no further increase in $q_{\rm peak}$ as we enter in the bubble phase at $\rho\sim0.4\rho_0$ (see the 
snapshots in \ref{fig:snapshots_xp50_T0}). This behavior was also seen in earlier calculations \citep{Watanabe03b, Nandi16}.
We find that the values of $S(q)$ are always slightly higher for neutrons than that of protons. This happens because of the Coulomb interaction that
increases the average distance between protons. Therefore, protons act less coherently than neutrons resulting in lower values of $S(q)$. We also observe that 
at $\rho\gtrsim 0.2\rho_0$, where we have irregular pasta phases (Fig. \ref{fig:snapshots_xp50_T0}), the shapes of $S(q)$ are not very regular unlike in
\citet{Horowitz08}. 

\begin{figure}
 \begin{center}
 \includegraphics[width=0.7\textwidth,angle=-90]{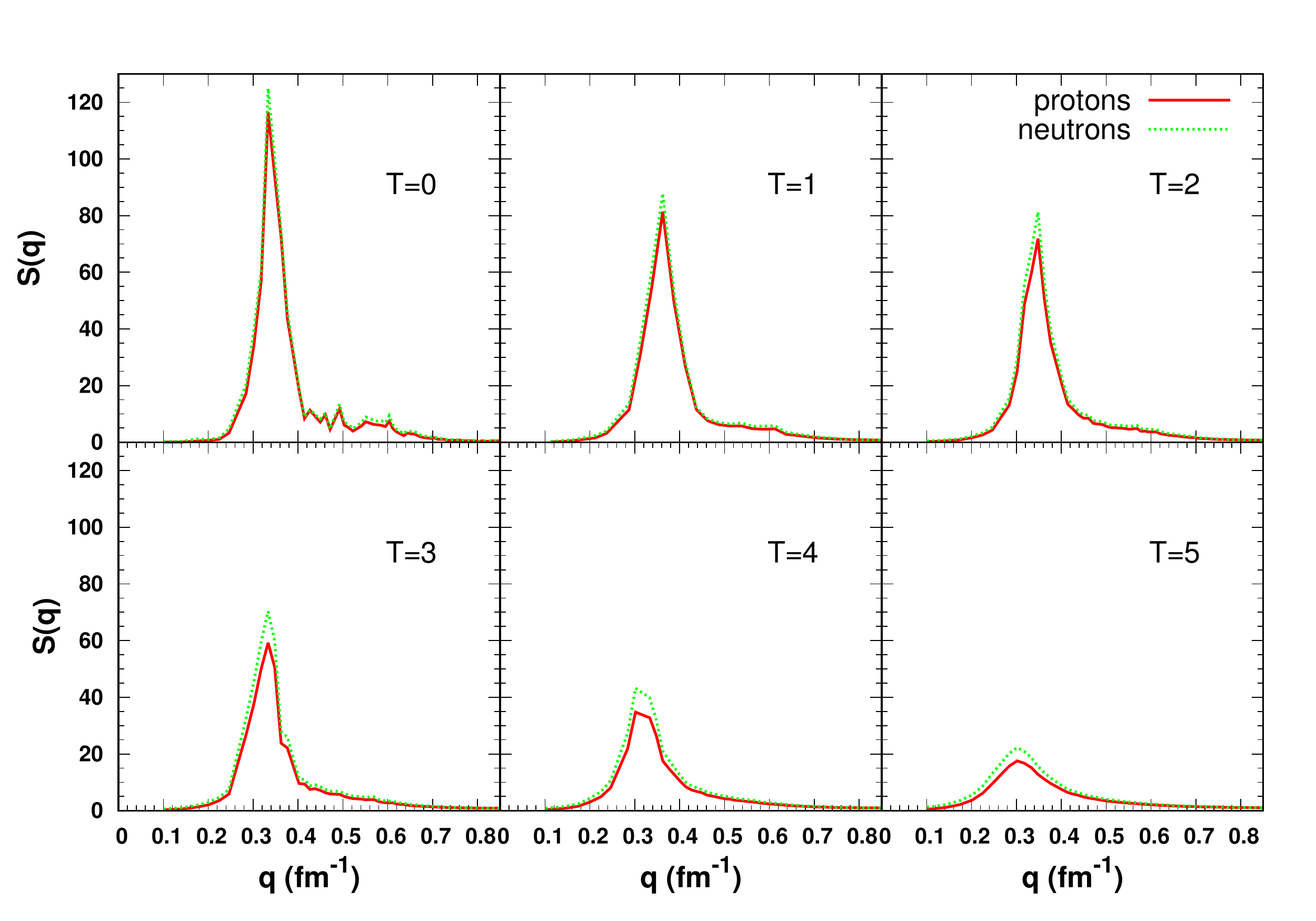}
 \caption{Static structure factor vs momentum transfer ($q$) for protons (solid line) and neutrons (dashed line) at $\rho=0.1\rho_0,\,Y_p=0.5$ and
 different temperatures. }
 \label{fig:sq_p100nb_xp50_allT}
 \end{center}
\end{figure}
\begin{figure}
 \begin{tabular}{cccccc}
  \includegraphics[width=0.32\textwidth]{snapshot_p100nb_xp50_T0_pix0000.jpg}&
  \includegraphics[width=0.32\textwidth]{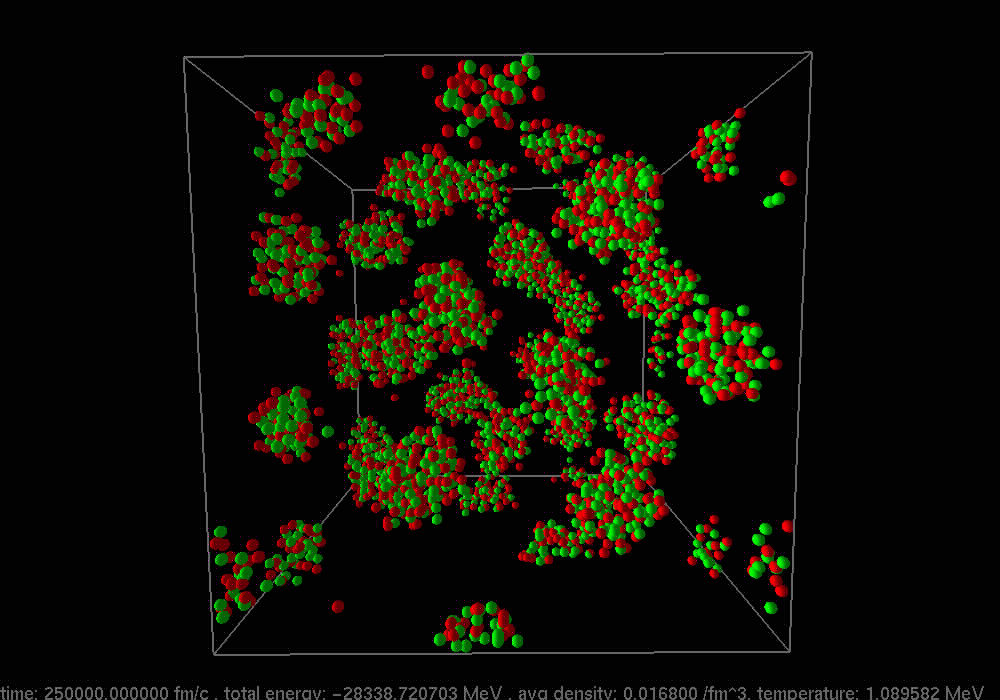}&
  \includegraphics[width=0.32\textwidth]{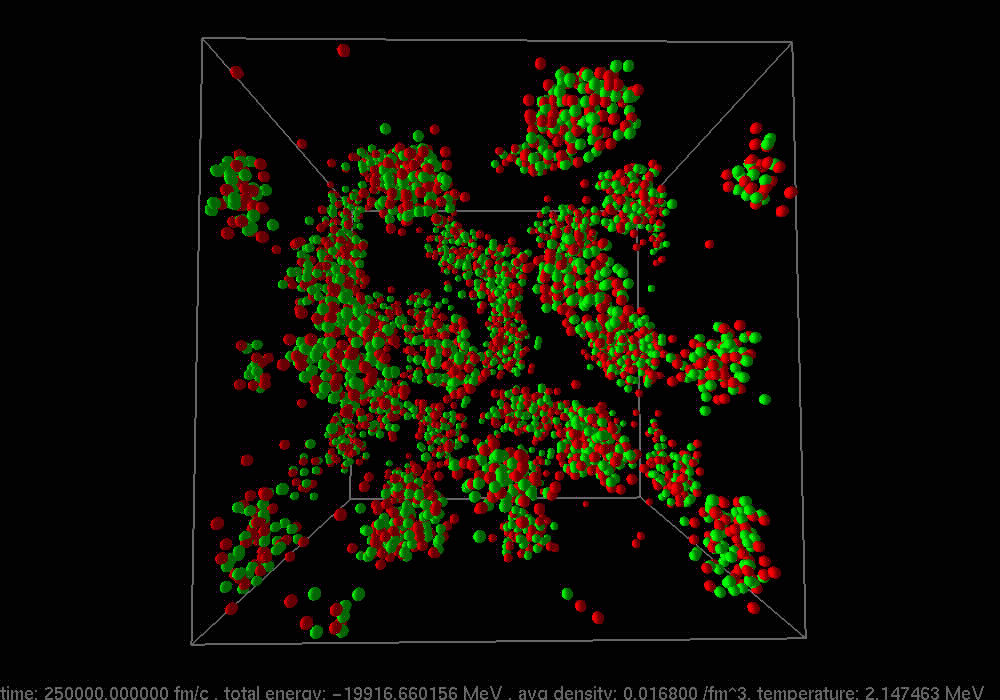}\\
  \includegraphics[width=0.32\textwidth]{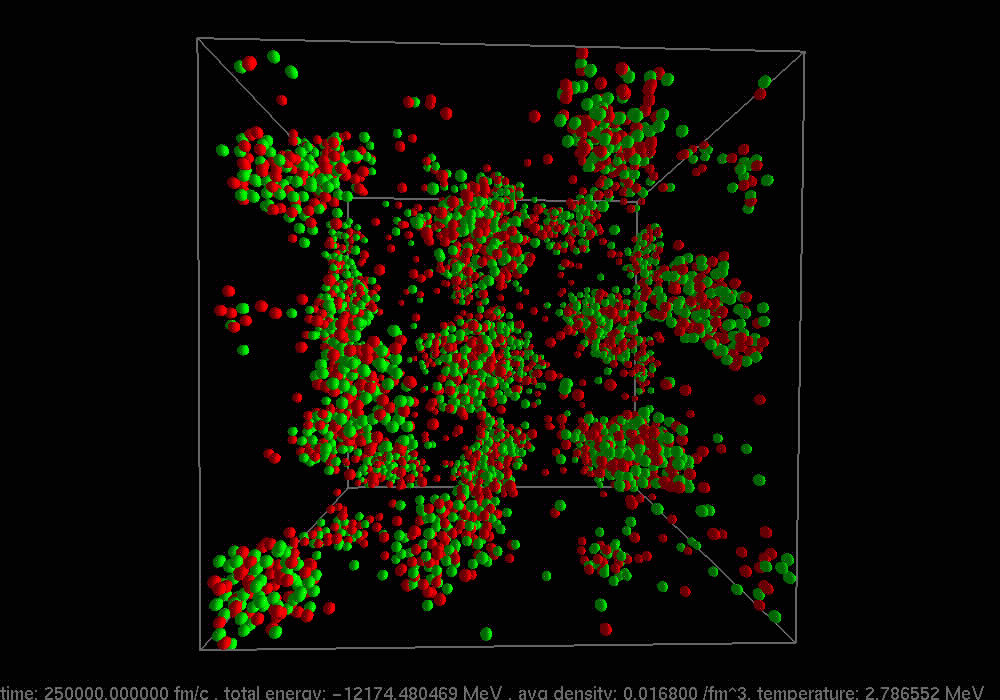}&
  \includegraphics[width=0.32\textwidth]{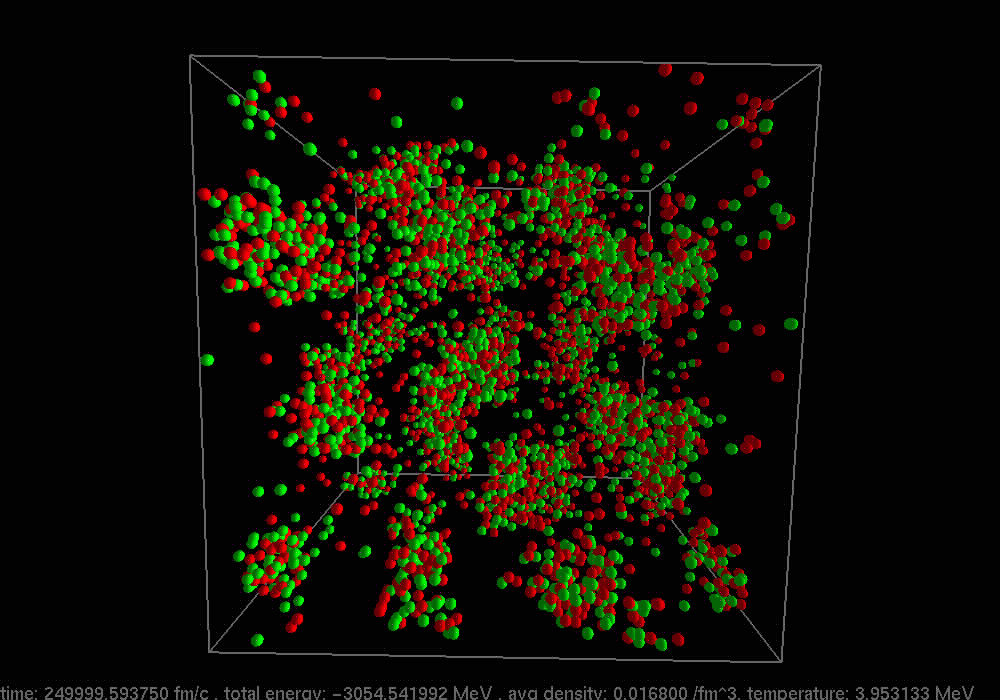}&
  \includegraphics[width=0.32\textwidth]{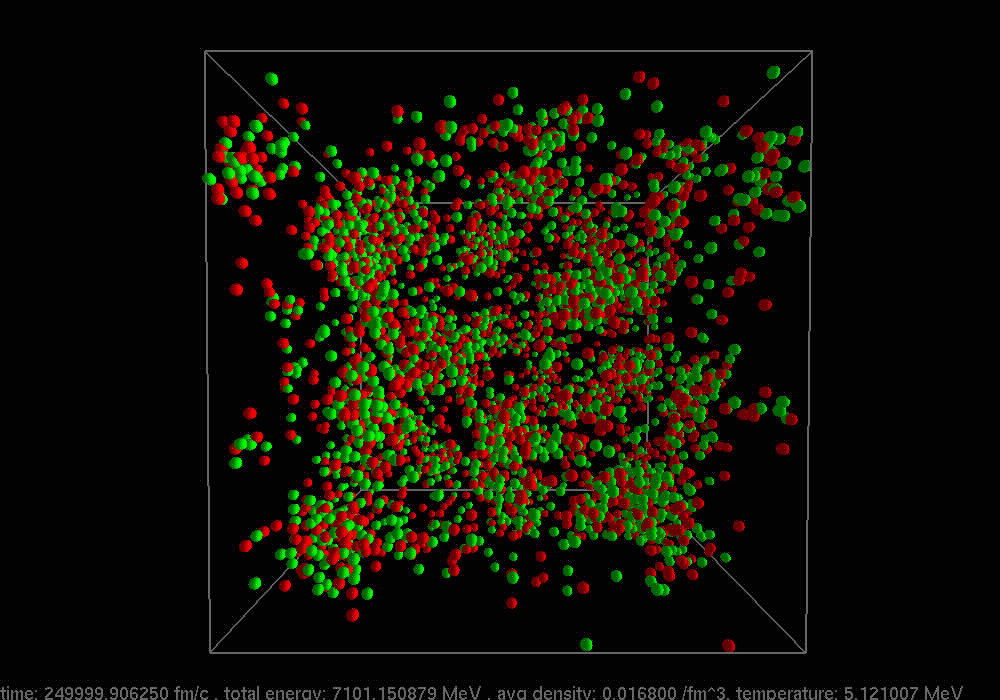}\\
 \end{tabular}
\caption{(Color online) simulation snapshots at $\rho=0.1\rho_0$, $Y_p=0.5$ and $T=0-5$ MeV. Here we use 8192 particles that corresponds to a box size of 62.47 fm.}
\label{fig:snapshots_p100nb_xp50}
\end{figure}

In Fig. \ref{fig:sq_p100nb_xp50_allT}, we plot $S(q)$ at $\rho=0.1\rho_0$, $Y_p=0.5$ and $T=0-5$ MeV. It can be observed that $S(q_{\rm peak})$ decreases with $T$.
As the temperature increases, more and more nucleons evaporate from clusters making the system increasingly uniform as can be seen in the snapshots 
(Fig. \ref{fig:snapshots_p100nb_xp50}). The average size of the clusters also decreases due to the presence of increasing number of smaller clusters.
These result in lower values of $S(q_{\rm peak})$ with $T$.

\begin{figure}
 \begin{tabular}{cccccc}
  \includegraphics[width=0.32\textwidth]{snapshot_p200nb_xp50_T0_pix0000.jpg}&
  \includegraphics[width=0.32\textwidth]{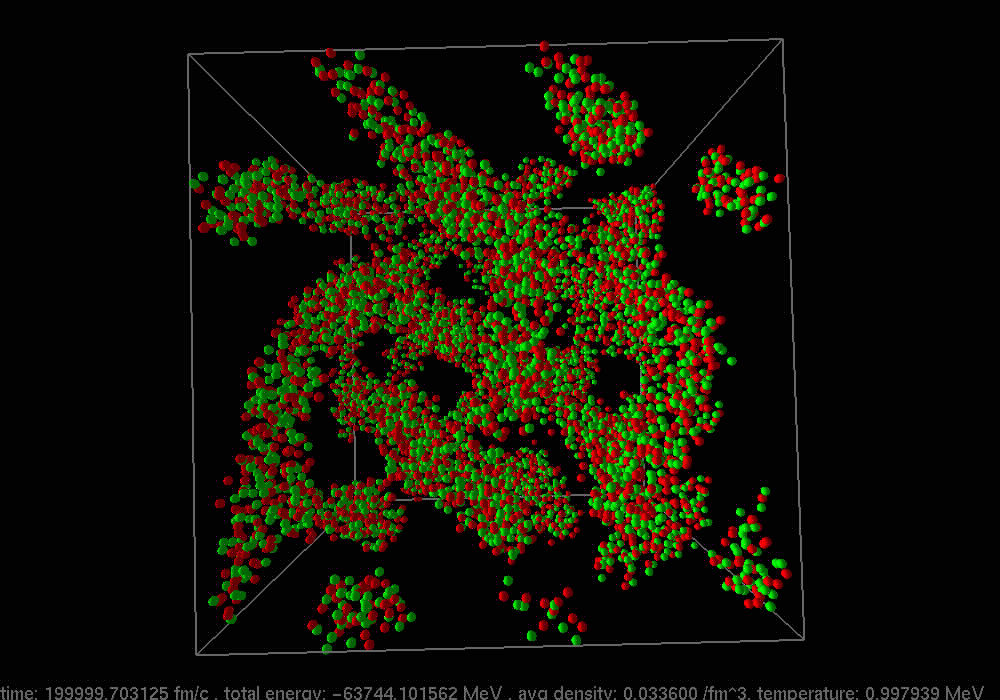}&
  \includegraphics[width=0.32\textwidth]{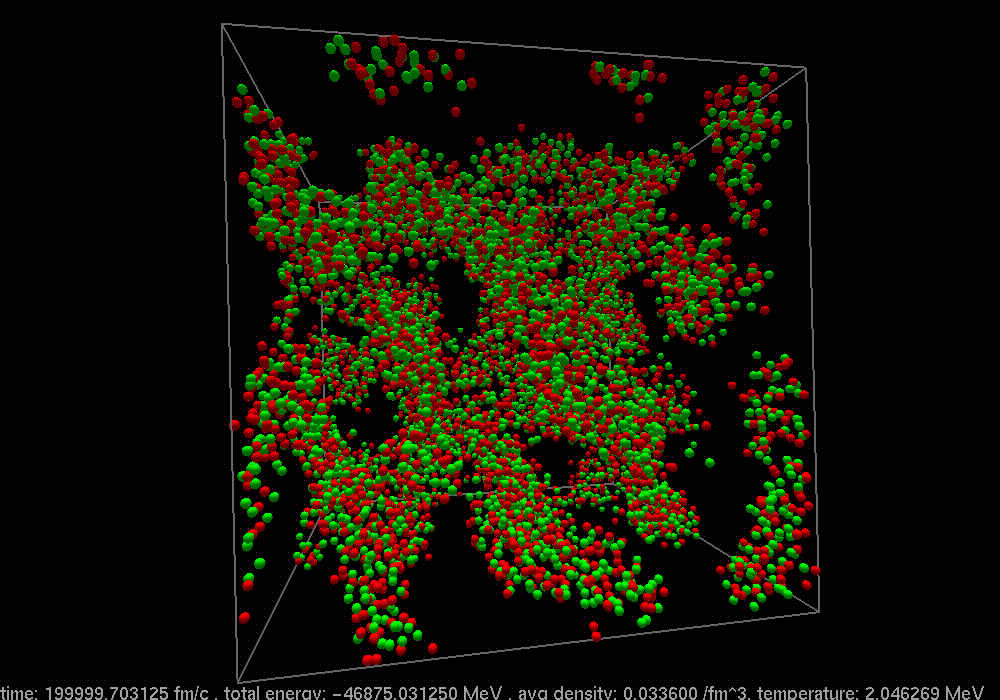}\\
  \includegraphics[width=0.32\textwidth]{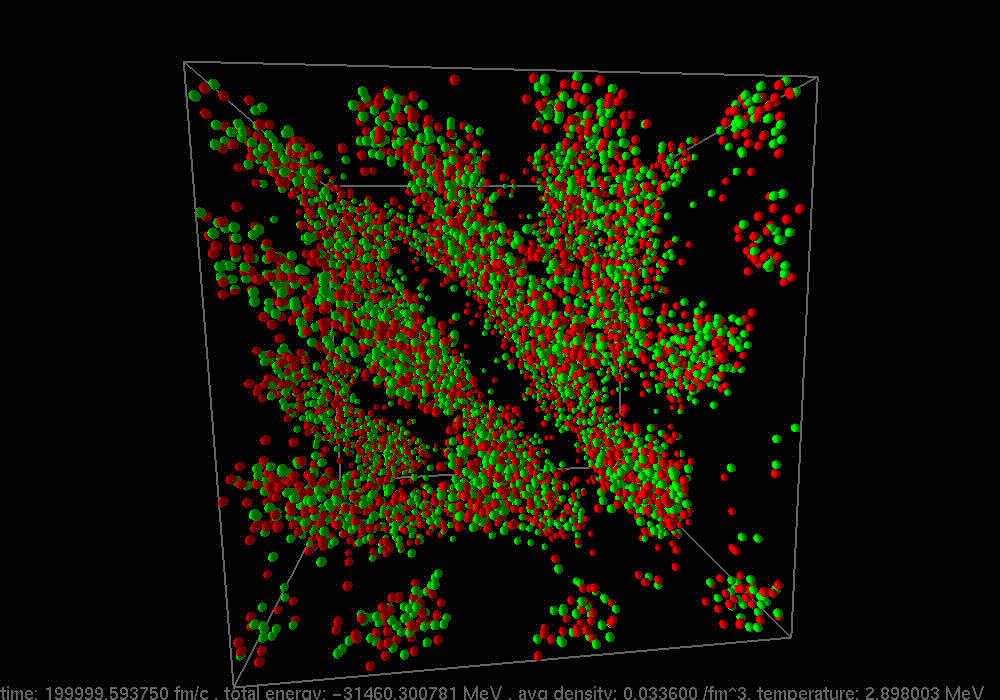}&
  \includegraphics[width=0.32\textwidth]{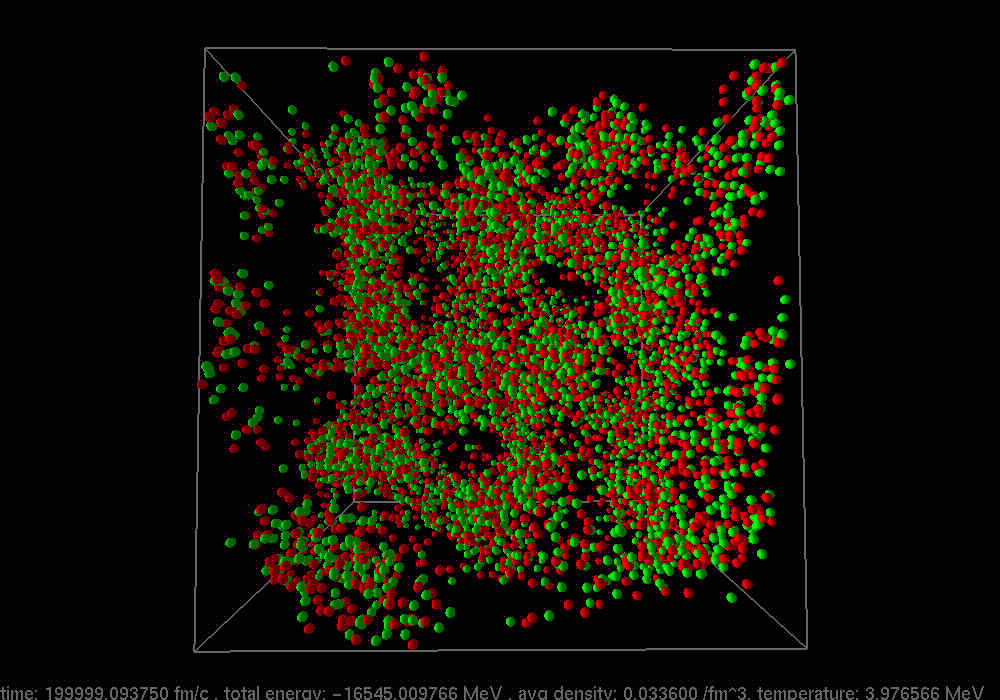}&
  \includegraphics[width=0.32\textwidth]{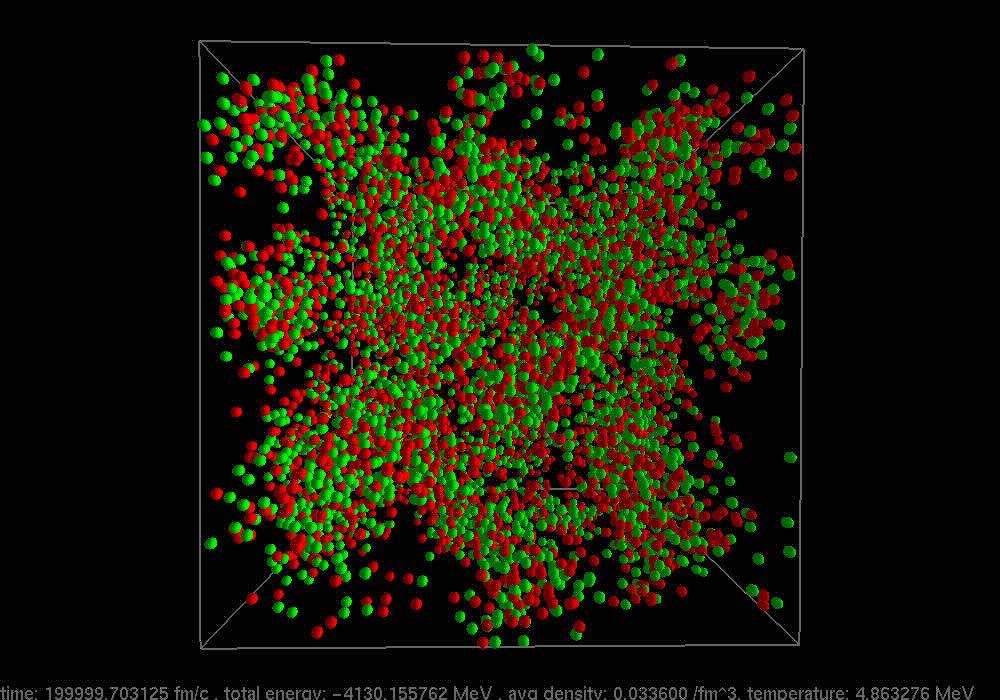}\\
 \end{tabular}
\caption{(Color online) simulation snapshots at $\rho=0.2\rho_0$, $Y_p=0.5$ and $T=0-5$ MeV. Here we use 8192 particles that corresponds to box size of 62.47 fm.}
\label{fig:snapshots_p200nb_xp50}
\end{figure}

An interesting behavior of $S(q)$ can be observed when we look at the $T$ dependence of pasta phases at $\rho\gtrsim0.2\rho_0$.
\begin{figure}
 \begin{center}
 \includegraphics[width=0.7\textwidth,angle=-90]{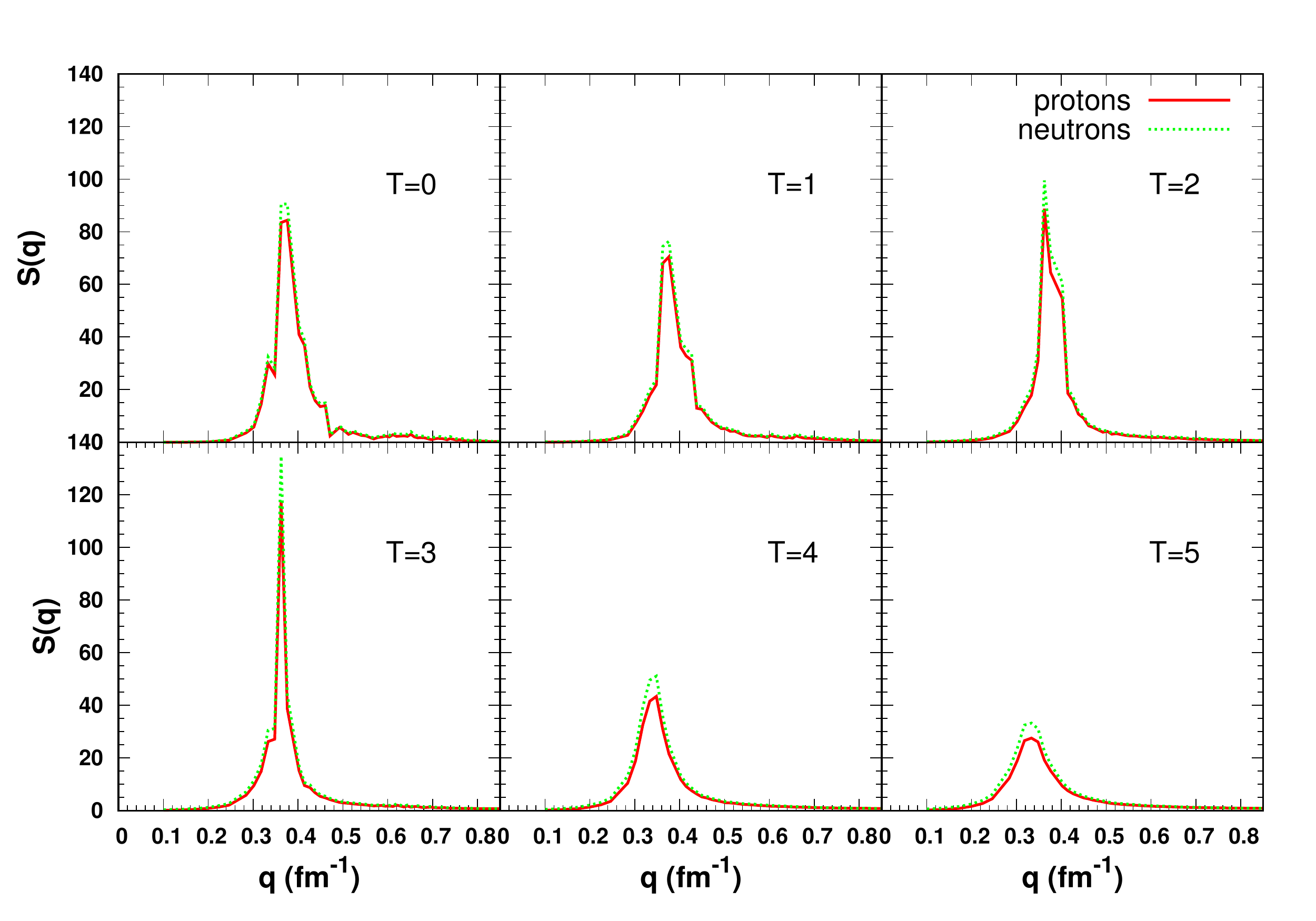}
 \caption{Static structure factor vs momentum transfer ($q$) for protons (solid line) and neutrons (dashed line) at $\rho=0.2\rho_0$, $Y_p=0.5$ and
 $T=0-5$ MeV. }
 \label{fig:sq_p200nb_xp50_allT}
 \end{center}
\end{figure}
In Fig. \ref{fig:sq_p200nb_xp50_allT} we plot $S(q)$ for $\rho=0.2\rho_0$ and various $T$ ($0-5$ MeV) for symmetric nuclear matter. 
With increasing $T$ we find that initially $S(q_{\rm peak})$ decreases but at $T=2$ it begins to rise reaching a very
high peak at $T=3$ after which it decreases gradually. This behavior of the $S(q)$ peak can be understood if we look at the corresponding snapshots shown in
Fig. \ref{fig:snapshots_p200nb_xp50}.
The snapshot at $T=0$ shows twisted cylinders with well-defined surfaces. At $T=1$ some of the bigger clusters get fragmented to form smaller clusters so that
the average size of the clusters decreases. As a result, the peak height of $S(q)$ gets reduced. But at $T=2$ clusters get so diffused that they begin to merge
and at $T=3$ most of the nucleons are connected to form a single big cluster (connected slab) giving rise to the very high peak. With further increase in temperature 
the matter becomes more uniform to give lower values of $S(q)$. 

\begin{figure}
\centering
 \includegraphics[width=0.7\textwidth,angle=-90]{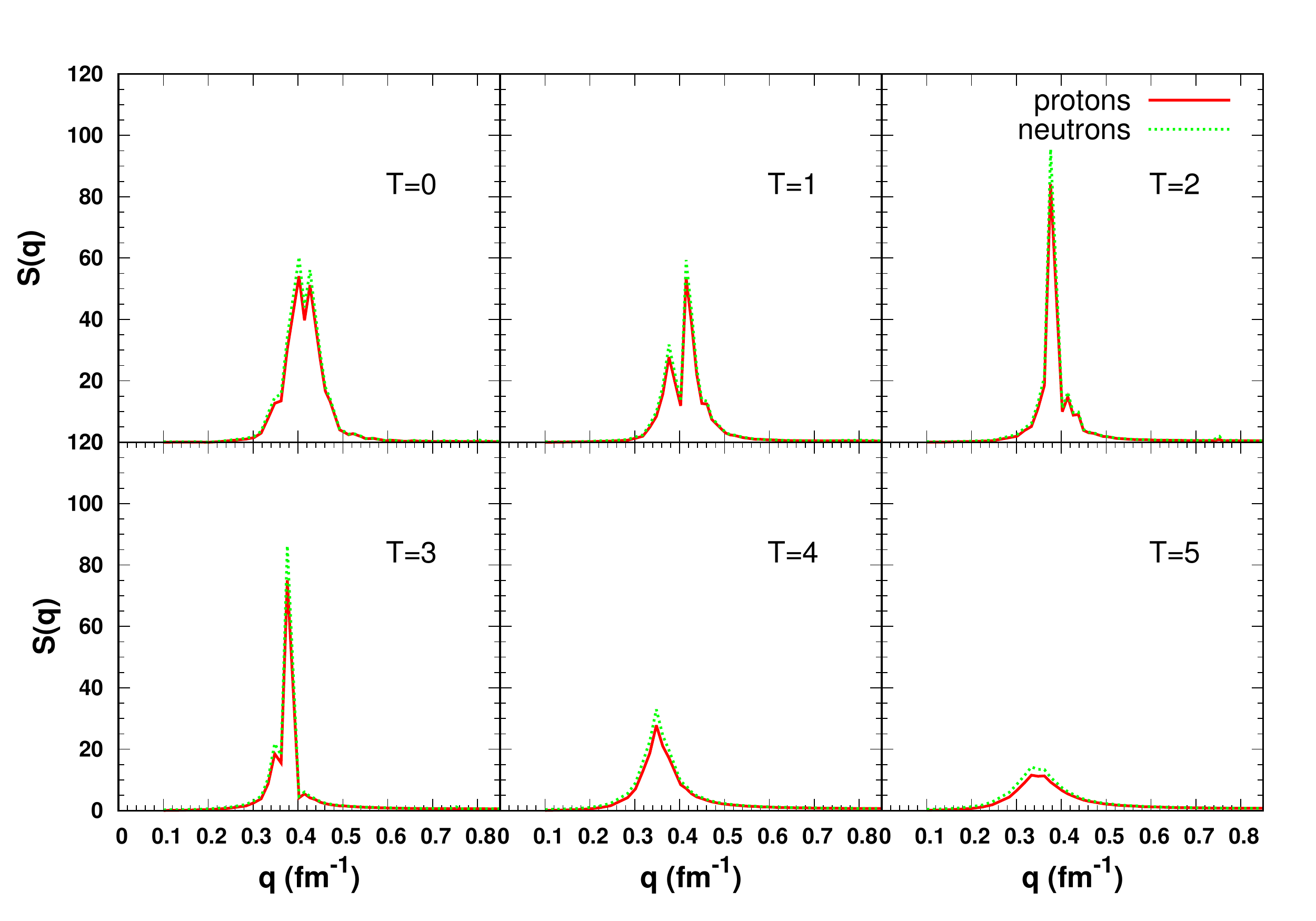}
  \caption{Same as Fig. \ref{fig:sq_p200nb_xp50_allT} but for $\rho=0.4\rho_0$.}
 \label{fig:sq_p400nb_xp50_allT}
\end{figure}

\begin{figure}
 \begin{center}
  \begin{tabular}{cccc}
   \includegraphics[width=0.24\textwidth]{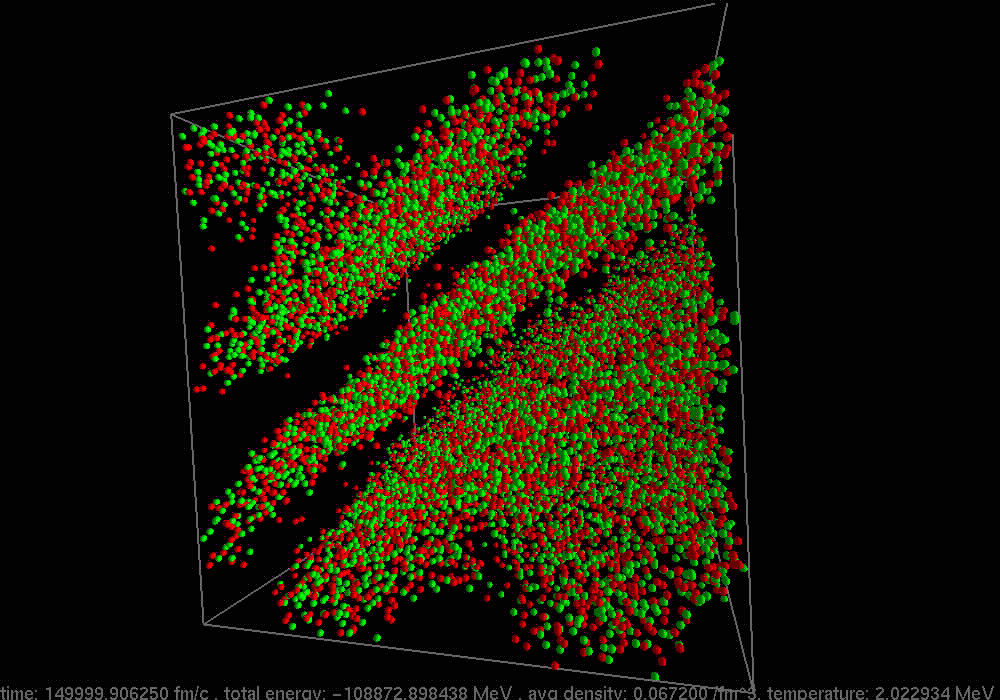}&
   \includegraphics[width=0.24\textwidth]{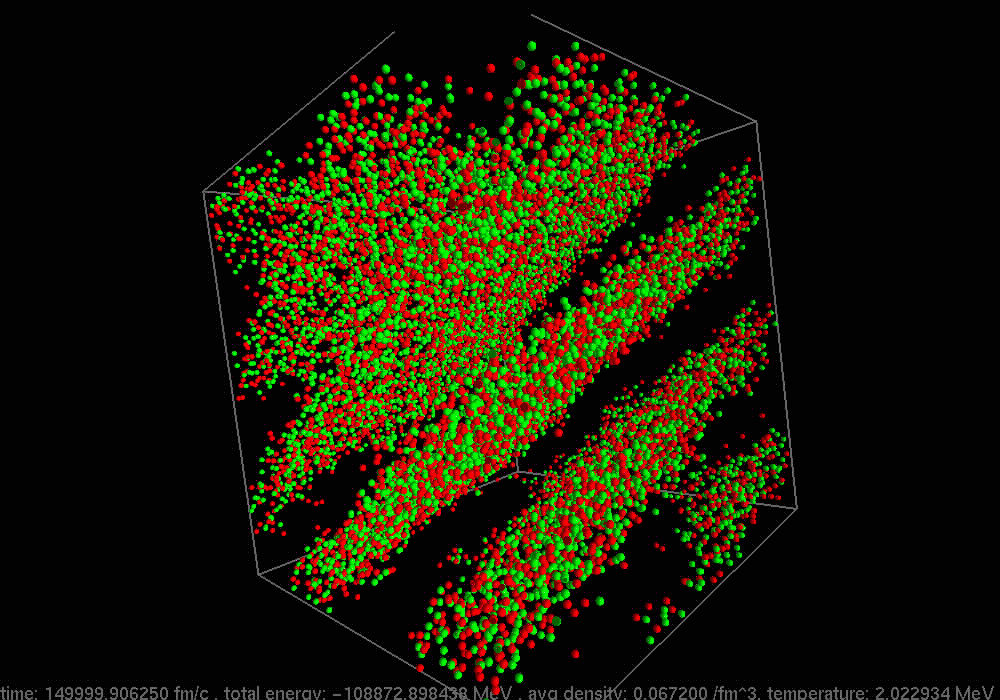}&
   \includegraphics[width=0.24\textwidth]{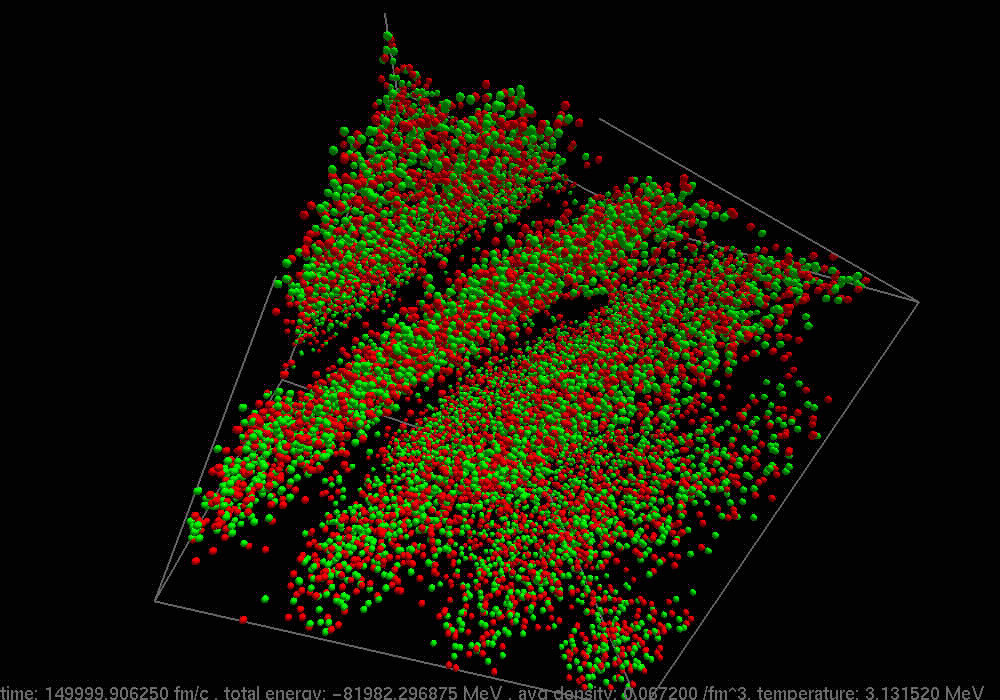}&
   \includegraphics[width=0.24\textwidth]{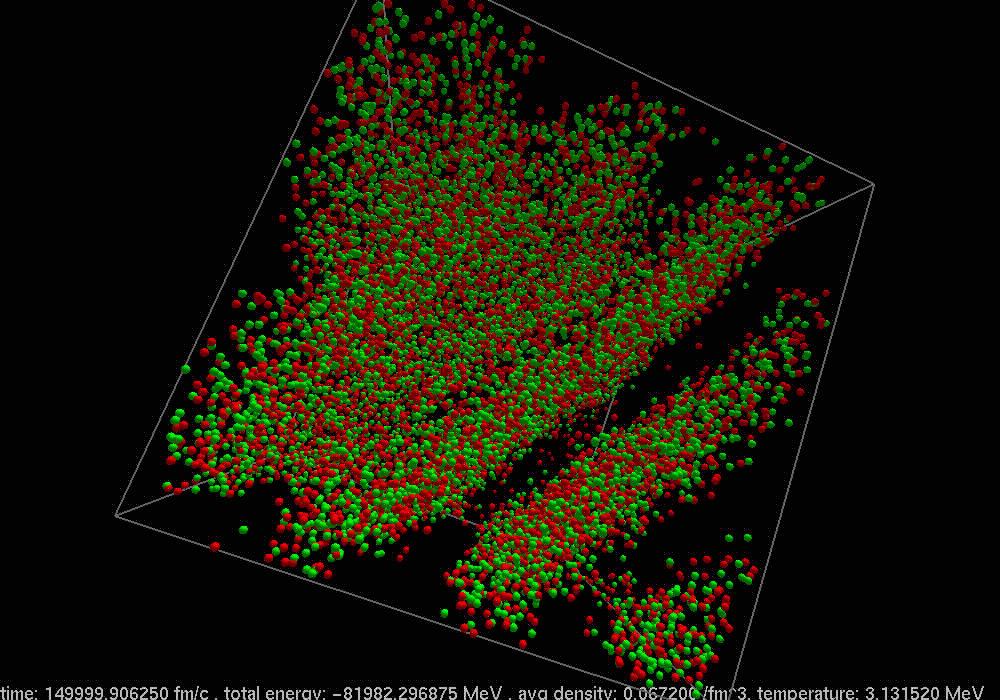}\\
  \end{tabular}
  \caption{Snapshots from simulations at $\rho=0.4\rho_0$, $T=2$ MeV (left 2 panels) and $T=3$ MeV (right 2 panels). Here we use 16384 nucleons.}
  \label{fig:snapshots_p400nb_xp50}
 \end{center}
\end{figure}

We also show $S(q)$ and relevant snapshots at $\rho=0.4\rho_0$ in Fig. \ref{fig:sq_p400nb_xp50_allT} and
Fig. \ref{fig:snapshots_p400nb_xp50}, respectively. As in $\rho=0.2\rho_0$, we find high and sharp peaks at $T=2$ and $3$ MeV. Snapshots in 
Fig. \ref{fig:snapshots_p400nb_xp50} reveal that this is the consequences of obtaining equidistant slabs at these temperatures.

\begin{table}
 \caption{Structure factors and Coulomb logarithms for symmetric nuclear matter ($Y_p=0.5$)}
  \begin{tabular}{cccccc|cccccc}\hline
    $T$       & $q_{\rm peak}$ & \multicolumn{2}{c}{$S(q_{\rm peak})$}&$\Lambda_{ep}^\eta$& $\Lambda_{ep}^\kappa$ & $T$ & $q_{\rm peak}$ & \multicolumn{2}{c}{$S(q_{\rm peak})$}&$\Lambda_{ep}^\eta$& $\Lambda_{ep}^\kappa$\\
    (MeV)     &   (fm$^{-1}$)  & protons  &neutrons                    &                   &                       &(MeV)&   (fm$^{-1}$)  & protons  & neutrons                  &                   &\\  
    \hline 
    \multicolumn{6}{c|}{$\rho=0.1\rho_0$}&\multicolumn{6}{c}{$\rho=0.2\rho_0$}\\ \hline
    0        & 0.348          &116.60    &124.88 &  16.61 &  18.06    & 0         & 0.370          & 83.98    & 90.51 & 11.59 & 12.31 \\
    1        & 0.363          & 81.41    & 87.89 &  13.39 &  14.63    & 1         & 0.370          & 69.26    & 75.43 & 10.11 & 10.76\\
    2        & 0.348          & 71.89    & 81.27 &  13.02 &  14.15    & 2         & 0.363          & 88.76    & 99.57 & 10.98 & 11.65\\ 
    3        & 0.334          & 59.23    & 70.27 &  12.78 &  13.77    & 3         & 0.363          & 117.74   &134.71 &  9.45 &  9.99\\
    4        & 0.311          & 34.24    & 42.33 &  10.24 &  11.01    & 4         & 0.348          &  43.42   & 51.12 &  9.34 &  9.82\\
    5        & 0.303          & 17.56    & 22.22 &   7.50 &   8.04    & 5         & 0.334          &  27.58   & 33.24 &  7.85 &  8.25\\
    \hline
    \multicolumn{6}{c|}{$\rho=0.3\rho_0$}&\multicolumn{6}{c}{$\rho=0.4\rho_0$}\\ \hline
    0        & 0.415          &  78.28   & 84.86 &  8.65  &  9.10     & 0         & 0.415          &  52.68   & 58.27 & 7.85 & 8.19\\
    1        & 0.396          &  55.70   & 61.65 &  7.51  &  7.90     & 1         & 0.427          &  38.60   & 42.55 & 5.33 & 5.57\\
    2        & 0.402          &  64.26   & 70.98 &  9.33  &  9.78     & 2         & 0.376          &  84.22   & 95.55 & 6.13 & 6.37\\
    3        & 0.363          &  68.64   & 79.11 &  8.02  &  8.37     & 3         & 0.376          &  75.07   & 86.29 & 5.59 & 5.80\\
    4        & 0.348          &  56.16   & 66.13 &  7.89  &  8.22     & 4         & 0.348          &  27.77   & 32.96 & 5.32 & 5.51\\
    5        & 0.348          &  25.58   & 30.49 &  6.09  &  6.34     & 5         & 0.334          &  11.57   & 14.24 & 3.84 & 3.98\\
    \hline
    \multicolumn{6}{c|}{$\rho=0.5\rho_0$}&\multicolumn{6}{c}{$\rho=0.6\rho_0$}\\ \hline
    0        & 0.415          &  49.39   & 55.30 &  4.84  & 5.04      & 0         & 0.415          & 30.48    &  33.31 & 3.84 & 3.96 \\
    1        & 0.376          &  31.03   & 35.71 &  5.43  & 5.62      & 1         & 0.376          & 19.09    &  22.19 & 3.15 & 3.25\\
    2        & 0.370          &  32.62   & 37.58 &  5.17  & 5.34      & 2         & 0.376          & 15.27    &  17.83 & 2.78 & 2.86\\
    3        & 0.363          &  25.96   & 30.48 &  4.15  & 4.29      & 3         & 0.348          &  4.79    &   5.98 & 1.71 & 1.76\\
    4        & 0.348          &  10.37   & 12.65 &  2.88  & 2.97      & 4         &                &          &        & 0.77 & 0.80\\
    5        &                &          &       &  1.79  & 1.85      & 5         &                &          &        & 0.59 & 0.62\\ 
        \hline\hline                     
  \end{tabular}
  \label{tab:peak_xp50}
\end{table}
We calculate $S(q)$ for all the densities and temperatures in the range $0.1-0.6\rho_0$ and $0-5$ MeV, respectively in similar fashion. 
In Table \ref{tab:peak_xp50} we compile the values of $q_{\rm peak}$  and $S(q_{\rm peak})$ for protons and neutrons for symmetric nuclear matter. 
In few cases we obtain double peaks (e.g. See the plot of $S(q)$ for $\rho=0.4\rho_0$ in Fig. \ref{fig:sq_allnb_xp50_T0}). We take average values
for both $q_{\rm peak}$  and $S(q_{\rm peak})$ in these cases. There are no results at $T=5$ MeV, $\rho=0.5\rho_0$ and $T=4-5$ MeV, $\rho=0.6\rho_0$ as
the matter becomes uniform at these conditions and therefore do not show any peak in $S(q)$. From the table we see that $S(q_{\rm peak})$ shows similar 
behavior as in $\rho/\rho_0=0.2$ and 0.4, discussed in previous paragraphs for other densities $\rho=0.3-0.5\rho_0$ also. Another interesting feature of the results is that
the value of $q_{\rm peak}$ generally decreases with temperature due to the presence of an increasing number of smaller clusters. 

\begin{figure}
 \begin{tabular}{cccccc}
  \includegraphics[width=0.32\textwidth]{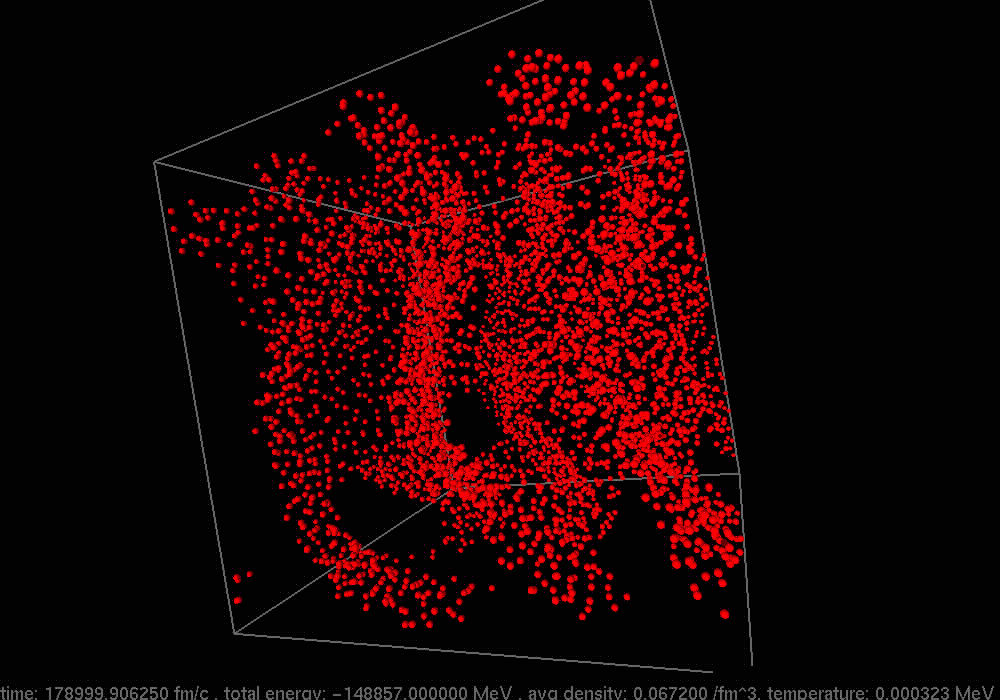}&
  \includegraphics[width=0.32\textwidth]{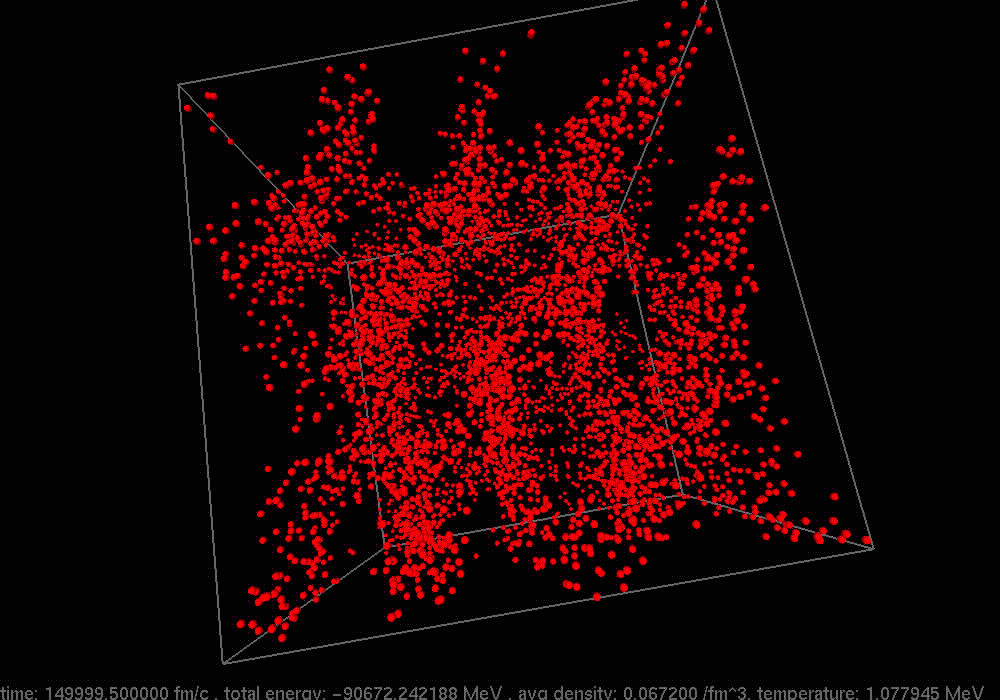}&
  \includegraphics[width=0.32\textwidth]{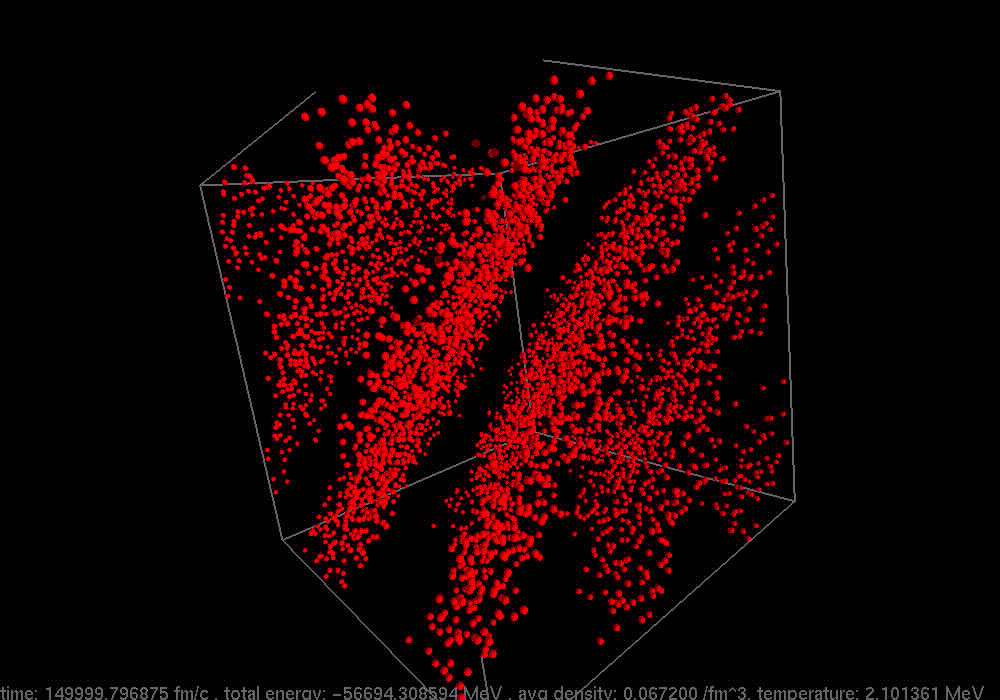}\\
  \includegraphics[width=0.32\textwidth]{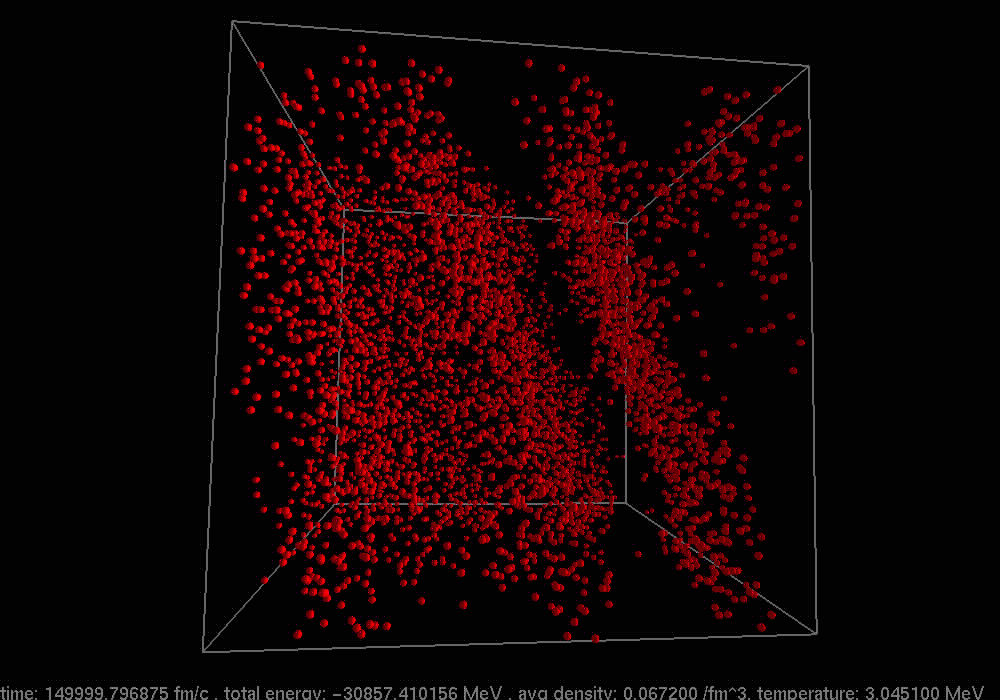}&
  \includegraphics[width=0.32\textwidth]{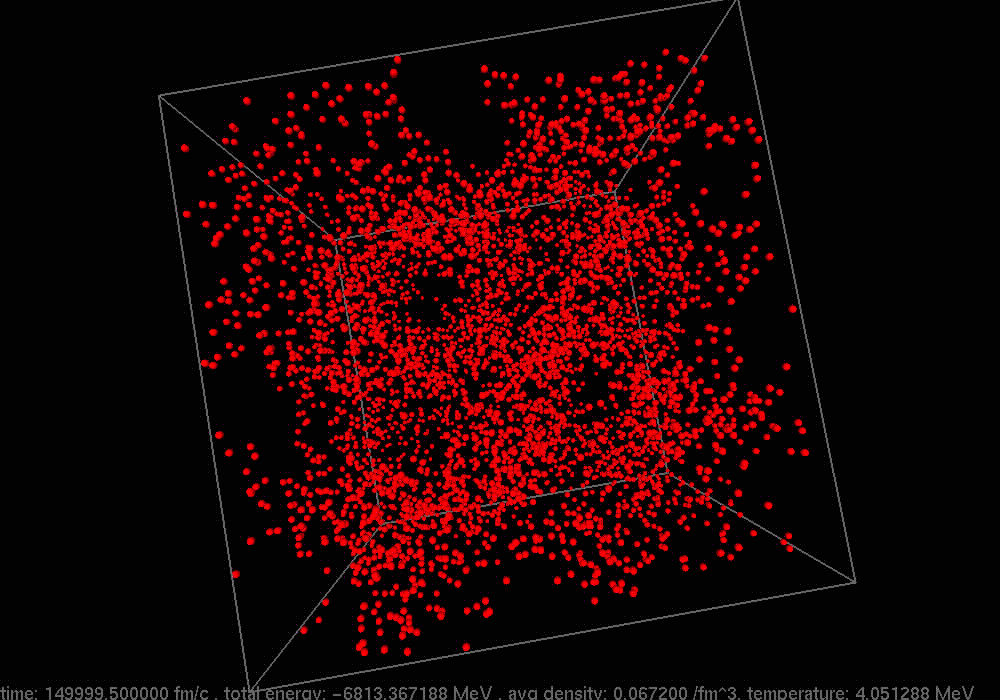}&
  \includegraphics[width=0.32\textwidth]{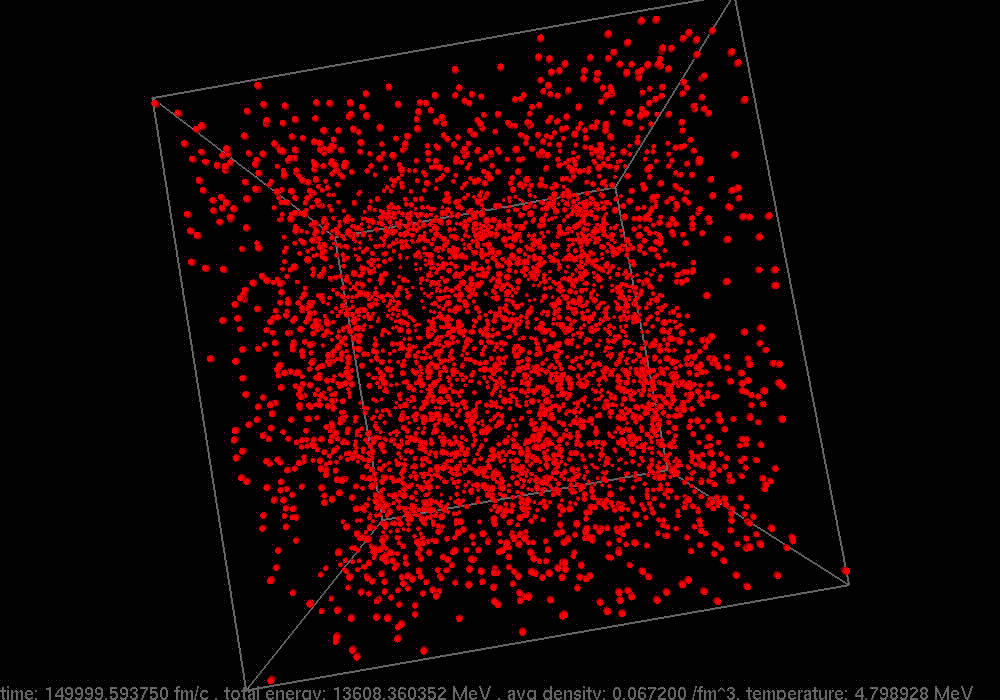}\\
 \end{tabular}
\caption{(Color online) Simulation snapshots of proton distributions at $\rho=0.4\rho_0$, $Y_p=0.3$ and $T=0-5$ MeV.
Here, 16384 nucleons (4864 protons) are used.}
\label{fig:snapshots_p400nb_xp30}
\end{figure}

Next, we calculate static structure factors for asymmetric nuclear matter with $Y_p=0.3$, relevant for supernova environment. 
\begin{figure}
 \begin{center}
 \includegraphics[width=0.7\textwidth,angle=-90]{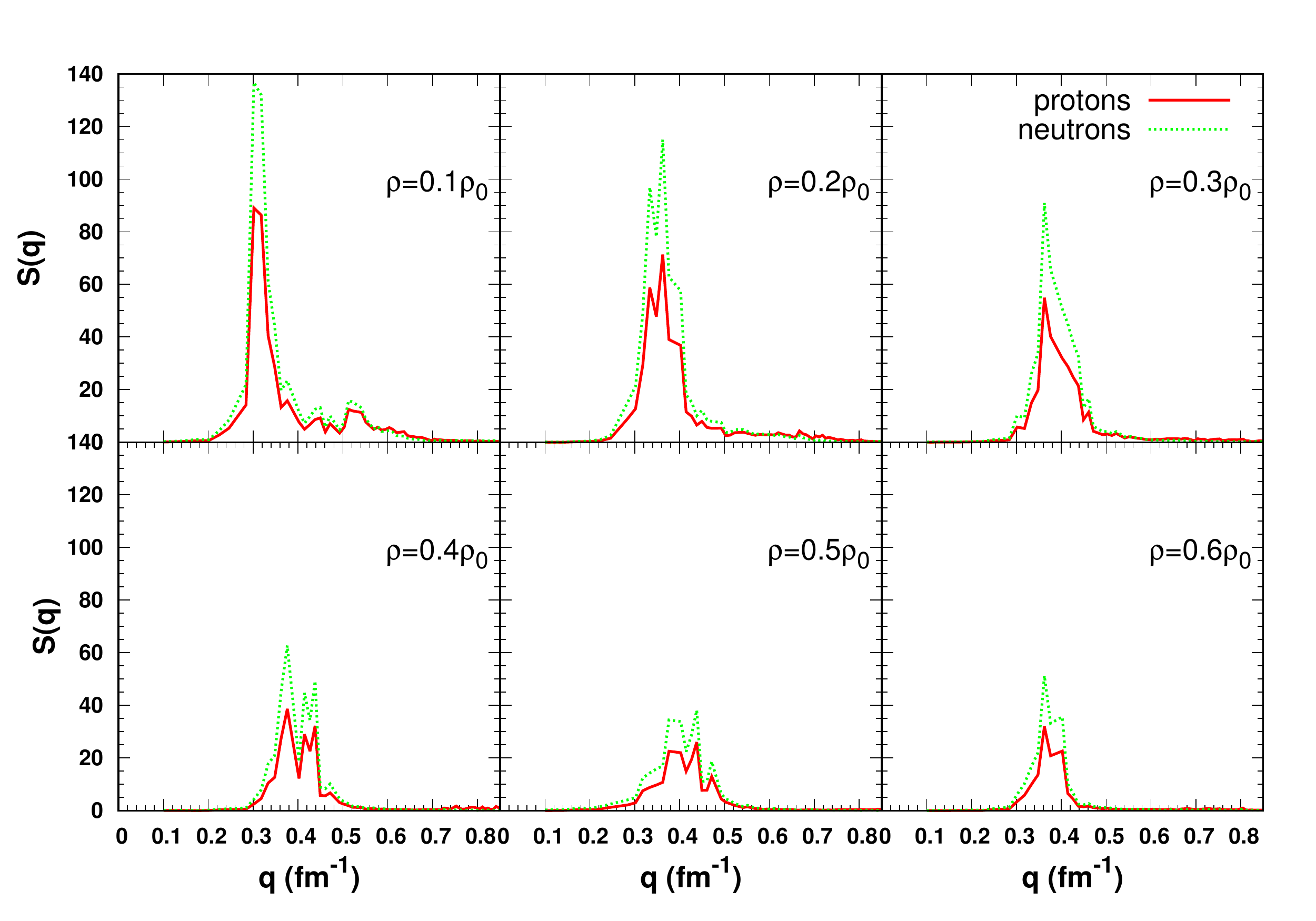}
 \caption{Static structure factor vs momentum transfer ($q$) for protons (solid line) and neutrons (dashed line) at $T=0$,  $Y_p=0.3$ and $\rho=0.1-0.6\rho_0$. }
 \label{fig:sq_allnb_xp30_T0}
 \end{center}
\end{figure}
In Fig. \ref{fig:sq_allnb_xp30_T0} we show $S(q)$ for $T=0$ and density range $0.1-0.6\rho_0$. At all densities $S_n(q_{\rm peak})$ are found to be much
higher than $S_p(q_{\rm peak})$. This happens because the clusters are neutron rich for this asymmetric nuclear matter. Similar to the symmetric matter, here
also $S(q)$ decreases with density and becomes irregular at densities $\rho\gtrsim0.2\rho_0$, when the pasta phase starts to appear. Likewise in $Y_p=0.5$, we 
calculate $S(q)$ for all the densities and temperatures considered. The results are given in Table \ref{tab:peak_xp30}. 
\begin{table}
\caption{Structure factors and Coulomb logarithms for asymmetric nuclear matter ($Y_p=0.3$)}
  \begin{tabular}{cccccc|cccccc}\hline
    $T$       & $q_{\rm peak}$ & \multicolumn{2}{c}{$S(q_{\rm peak})$} & $\Lambda_{ep}^\eta$ &$\Lambda_{ep}^\kappa$ & $T$ & $q_{\rm peak}$ & \multicolumn{2}{c}{$S(q_{\rm peak})$}& $\Lambda_{ep}^\eta$ &$\Lambda_{ep}^\kappa$  \\
     (MeV)    &   (fm$^{-1}$)  & protons  &neutrons                    &                     &                      & (MeV)     &   (fm$^{-1}$)  & protons  & neutrons \\  
                \hline
                \multicolumn{6}{c|}{$\rho=0.1\rho_0$}&\multicolumn{6}{c}{$\rho=0.2\rho_0$}\\ \hline
     0        & 0.310          & 87.82    &134.42    &11.85 & 13.24    &   0         & 0.363          & 71.38    &115.08 &11.13 & 12.05\\
     1        & 0.318          & 48.40    & 69.90    &12.39 & 13.79    &   1         & 0.348          & 67.06    &104.85 &10.19 & 11.01\\
     2        & 0.302          & 60.35    & 80.69    &13.31 & 14.70    &   2         & 0.341          & 63.46    & 92.80 &10.87 & 11.65\\ 
     3        & 0.302          & 55.58    & 68.80    &12.77 & 14.04    &   3         & 0.318          & 50.47    & 70.88 &10.94 & 11.64\\
     4        & 0.284          & 38.57    & 45.55    &12.83 & 13.96    &   4         & 0.302          & 42.98    &  58.39&10.32 & 10.95\\
     5        & 0.284          & 17.44    & 20.05    & 8.76 &  9.50    &   5         & 0.284          & 27.53    &  36.72& 8.93 &  9.46\\
                \hline
                \multicolumn{6}{c|}{$\rho=0.3\rho_0$}&\multicolumn{6}{c}{$\rho=0.4\rho_0$}\\ \hline
     0         & 0.363          &  54.95   & 90.94   & 7.35 & 7.87     & 0           & 0.410          &  33.31   & 52.26 & 5.55 & 5.88\\
     1         & 0.376          &  47.72   & 73.69   & 7.70 & 8.23     & 1           & 0.402          &  58.52   & 87.93 & 6.58 & 6.96\\
     2         & 0.334          &  58.68   & 89.64   & 8.81 & 9.30     & 2           & 0.348          &  350.1   & 529.1 &12.98&13.59\\
     3         & 0.326          &  40.85   & 59.68   & 8.29 & 8.71     & 3           & 0.334          &  64.66   & 94.17 & 6.05 & 6.31 \\
     4         & 0.318          &  34.93   & 49.29   & 7.62 & 8.00     & 4           & 0.318          &  17.94   & 25.40 & 5.00 & 5.22\\
     5         & 0.302          &  17.95   & 24.74   & 6.14 & 6.44     & 5           & 0.318          &   8.95   & 12.09 & 3.90 & 4.08\\
                \hline
                \multicolumn{6}{c|}{$\rho=0.5\rho_0$}&\multicolumn{6}{c}{$\rho=0.6\rho_0$}\\ \hline
     0         & 0.389          &  22.30   & 34.15   & 5.11 & 5.37     & 0           & 0.363          & 32.01    &  51.33 & 4.05 & 4.21\\
     1         & 0.370          &  24.87   & 37.70   & 4.34 & 4.54     & 1           & 0.402          & 13.45    &  19.40 & 3.12 & 3.24\\
     2         & 0.348          &  19.19   & 28.65   & 4.55 & 4.73     & 2           & 0.334          &  8.75    &  12.69 & 2.33 & 2.41 \\
     3         & 0.333          &  12.40   & 17.90   & 3.70 & 3.85     & 3           & 0.318          &  1.83    &   2.36 & 1.14 & 1.19 \\
     4         &                &          &         & 2.50 & 2.61     & 4           &                &          &        & 0.66 & 0.70\\
     5         &                &          &         & 1.63 & 1.71     & 5           &                &          &        & 0.64 & 0.68\\
     
        \hline\hline                     
  \end{tabular}
  \label{tab:peak_xp30}
\end{table}
The general trend of $S(q)$ with density and temperature is similar to the case of symmetric nuclear matter. At $\rho=0.4\rho_0$, we find a surprisingly sharp and 
high peak at $T=2$ MeV. To investigate the cause for this behavior we look at the corresponding snapshots shown in Fig. \ref{fig:snapshots_p400nb_xp30}. We do not show the neutrons
to increase the visibility. At $T=0, 1$ MeV we find structures intermediate between cylinders and slabs. However, at $T=2$ we obtain almost perfect equidistant slabs
that give rise to the the sharp and high peak in $S(q)$. With further increase in $T$, $S(q_{\rm peak})$ decreases as the slabs slowly merge and form bubble phase 
at $T=4$ MeV, whereas at $T=5$ MeV we get almost uniform matter. 

We also calculate $S_p(q)$ and $S_n(q)$ for very asymmetric nuclear matter with $Y_p=0.1$, which is close to the value expected in the inner crust of neutron stars.
\begin{figure}
 \begin{center}
 \includegraphics[width=0.7\textwidth,angle=-90]{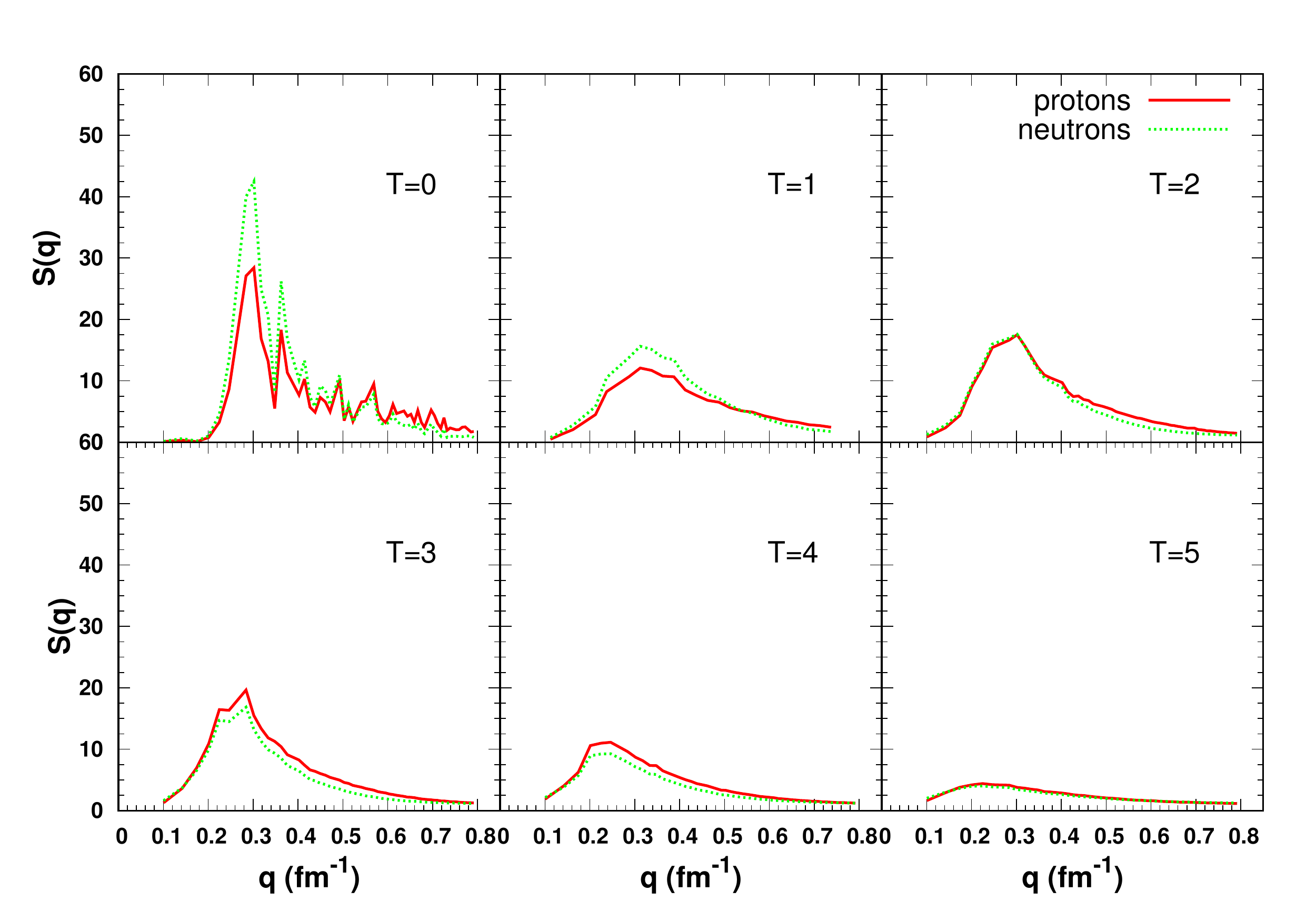}
 \caption{Static structure factor vs momentum transfer ($q$) for protons (solid line) and neutrons (dashed line) at $\rho=0.1\rho_0$,  $Y_p=0.1$ and $T=0-5$ MeV. }
 \label{fig:sq_p100nb_xp10_Tall}
 \end{center}
\end{figure}
In Fig. \ref{fig:sq_p100nb_xp10_Tall} we present $S(q)$ for neutrons and protons at $\rho=0.1\rho_0$ and $T=0-5$ MeV. At $T=0$, both $S_n(q)$ and $S_p(q)$ show
oscillatory behavior which likely indicates very irregular arrangement of clusters at this condition. Note that at $T\gtrsim 2$
MeV, $S_p(q)>S_n(q)$ when $q$ is not very small. This is because at $Y_p=0.1$ the clusters are very neutron rich and neutrons extend far beyond
the proton surface of the clusters leading to larger form factors for neutrons than protons. The form factor is more effective to reduce $S(q)$ at larger $q$ 
and hence results in smaller structure factors for neutrons than protons, at larger $q$ .
In Table \ref{tab:peak_xp10} we accumulate results for different $T$ and $\rho$ for $Y_p=0.1$. At this $Y_p$ we show results of $S(q)$ only for few values
of $\rho$ and $T$, because the phase diagram in $\rho-T$ plane is much smaller in this case \citep{Nandi17}.
\begin{table}
\caption{Structure factors and Coulomb logarithms for asymmetric nuclear matter ($Y_p=0.1$)}
  \begin{tabular}{cccccc|cccccc}\hline
    $T$       & $q_{\rm peak}$ & \multicolumn{2}{c}{$S(q_{\rm peak})$} &$\Lambda_{ep}^\eta$&$\Lambda_{ep}^\kappa$ & $T$ & $q_{\rm peak}$ & \multicolumn{2}{c}{$S(q_{\rm peak})$}&$\Lambda_{ep}^\eta$&$\Lambda_{ep}^\kappa$  \\
     (MeV)    &   (fm$^{-1}$)  & protons  &neutrons                     &                   &                      &(MeV)&   (fm$^{-1}$)  & protons  & neutrons                  &                   &\\  
                \hline
                 \multicolumn{6}{c|}{$\rho=0.1\rho_0$}&\multicolumn{6}{c}{$\rho=0.2\rho_0$}\\ \hline
      0        & 0.302          & 28.42    & 42.49  & 5.87 & 7.34  & 0         & 0.508          & 11.28    & 23.85 & 1.39 & 1.87\\
      1        & 0.348          & 12.33    & 15.89  & 5.13 & 6.32  & 1         & 0.461          & 12.51    & 21.54 & 1.98 & 2.50\\
      2        & 0.302          & 17.51    & 17.59  & 8.15 & 9.65  & 2         & 0.348          &  9.93    & 13.64 & 3.91 & 4.54\\ 
      3        & 0.284          & 19.64    & 16.84  & 9.31 &10.82  & 3         & 0.284          &  8.72    & 10.90 & 4.66 & 5.26\\
      4        & 0.246          & 11.12    &  9.26  & 7.10 & 8.12  & 4         & 0.284          &  6.55    &  7.56 & 4.19 & 4.67\\
      5        & 0.225          &  4.40    &  4.02  & 3.62 & 4.13  & 5         & 0.284          &  4.18    &  4.80 & 3.33 & 3.70\\
                \hline
                \multicolumn{6}{c|}{$\rho=0.3\rho_0$}&\multicolumn{6}{c}{$\rho=0.4\rho_0$}\\ \hline
      0        & 0.551          &  13.14   & 28.06  & 0.97 & 1.27  & 0         & 0.537          &  10.27   & 22.00 & 0.70 & 0.87 \\
      1        & 0.503          &   8.66   & 15.70  & 1.38 & 1.68  & 1         & 0.503          &   3.82   &  6.48 & 0.92 & 1.09\\
      2        & 0.415          &   5.31   &  7.89  & 2.18 & 2.50  & 2         &                &          &       & 1.22 & 1.38\\
      3        &                &          &        & 2.49 & 2.78  & 3         &                &          &       & 1.32 & 1.47\\
      4        &                &          &        & 2.38 & 2.63  & 4         &                &          &       & 1.33 & 1.47\\
      5        &                &          &        & 2.12 & 2.34  & 5         &                &          &       & 1.31 & 1.45\\
                 \hline
                \multicolumn{6}{c|}{$\rho=0.5\rho_0$}&\multicolumn{6}{c}{$\rho=0.6\rho_0$}\\ \hline
      0         & 0.542         &   4.58   &  9.19 &0.37  & 0.45 &  0          &                &          &       &0.02  & 0.02\\
      1         &               &          &       &0.43  & 0.50 &  1          &                &          &       &0.18  & 0.21\\ 
      2         &               &          &       &0.58  & 0.65 &  2          &                &          &       &0.33  & 0.37\\
      3         &               &          &       &0.70  & 0.78 &  3          &                &          &       &0.45  & 0.51\\
      4         &               &          &       &0.79  & 0.88 &  4          &                &          &       &0.55  & 0.61\\
      5         &               &          &       &0.87  & 0.96 &  5          &                &          &       &0.64  & 0.70\\     

         \hline\hline                     
  \end{tabular}
  \label{tab:peak_xp10}
\end{table}

\subsection{Transport coefficients}
In this section we calculate transport coefficients $\eta,\,\kappa$ and $\sigma$ from Eqs. (\ref{eq:tc}-\ref{eq:sv}) after determining the Coulomb logarithm in
Eqs. (\ref{eq:clc}-\ref{eq:clv}) using the results of $S_p(q)$ obtained in the previous subsection. The values of the Coulomb logarithms ($\Lambda_{ep}^\eta$ and
$\Lambda_{ep}^\kappa$) are given in the last two columns of Table \ref{tab:peak_xp50}-\ref{tab:peak_xp10}. It can be seen that generally $\Lambda$s decrease
with $T$ and $\rho$ at $Y_p=0.5$ and $0.3$. However, due the increase in $S(q_{\rm peak})$ at intermediate $T$, as discussed earlier, $\Lambda$s  increases at these
temperatures. At $Y_p=0.1$, Coulomb logarithms slowly increase with $T$ when the matter is uniform at larger densities ($\rho\gtrsim0.4\rho_0$).

In order to make $\Lambda$s readily available for future use we fit the data as following. We fit both the $\Lambda$s as a function of $\rho/\rho_0$ for a fixed $T$.
For $Y_p=0.5$ and $Y_p=0.3$ good fits are obtained if we choose
\begin{equation}
  \Lambda_{\eta,\kappa}^{\rm fit} = \sum_{i=0}^{4}a_i\,[ln(\rho/\rho_0)]^{-i}\, ,\, \textrm{for a fixed}\,\, T, \label{eq:fit_yp5}
\end{equation}
and for $Y_p=0.1$ we use
 \begin{equation}
     \Lambda_{\eta,\kappa}^{\rm fit} = \sum_{i=0}^{4}a_i\,[ln(\rho/\rho_0)]^i\, ,\, \textrm{for a fixed}\,\, T. \label{eq:fit_yp1}
\end{equation}
The fit parameters are given in the appendix. The maximum fitting residual ($|1-\Lambda_{\rm fit}/\Lambda|$) is $\lesssim 8\%$ except at $Y_p=0.3,
\rho=0.4\rho_0$ and $T=2$ MeV, where the residual is $\sim 22\%$ as $\Lambda$s rise suddenly at this point due to the presence of a slab phase. The calculation
of transport coefficients from $\Lambda$s are straightforward (See Eqs. (\ref{eq:tc}-\ref{eq:sv})).

\begin{figure}
  \includegraphics[width=0.65\textwidth, angle=-90]{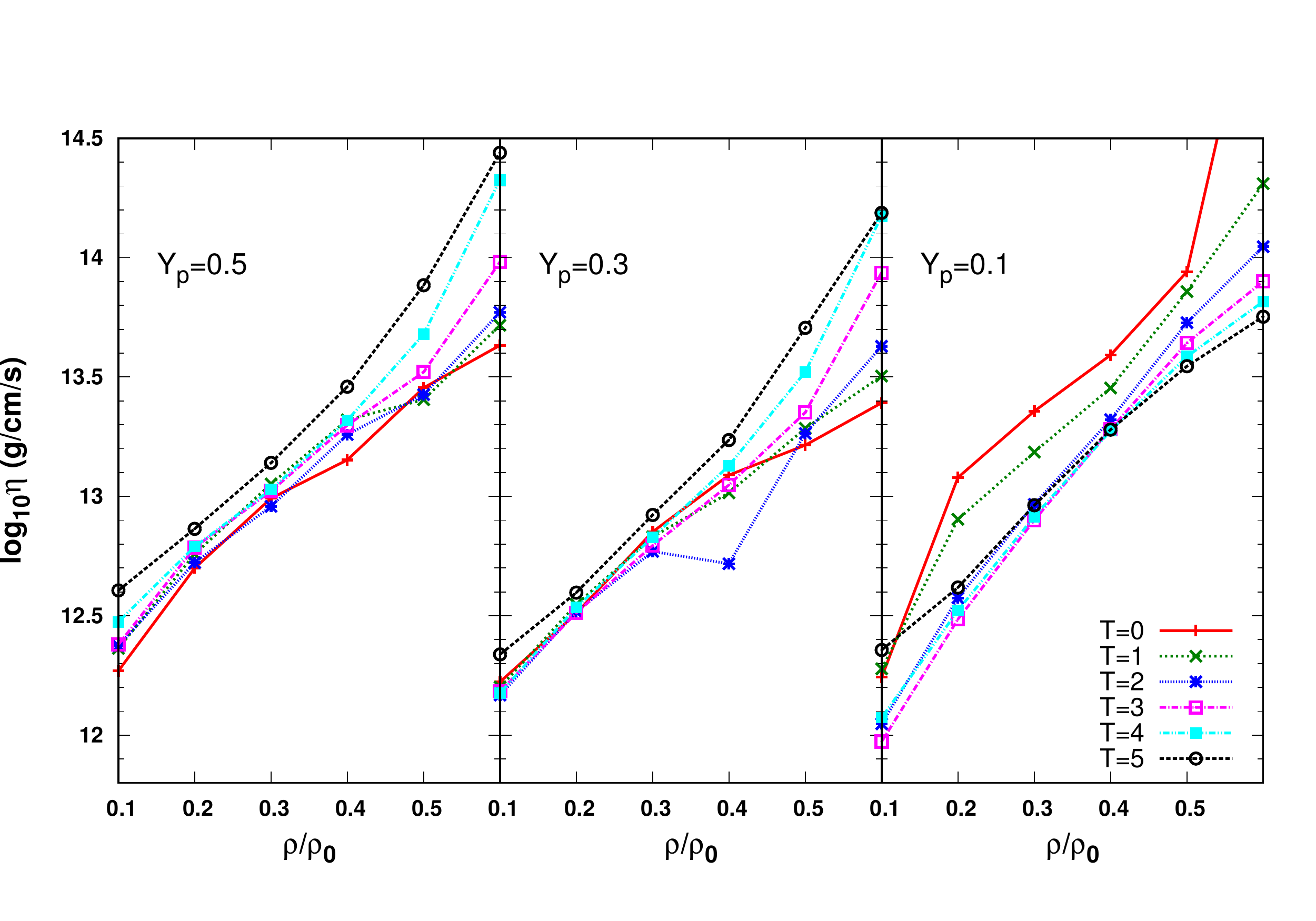}
 \caption{Shear viscosity as a function of density at different temperatures and proton fractions.}
 \label{fig:eta}
\end{figure}
\begin{figure}
  \includegraphics[width=0.65\textwidth, angle=-90]{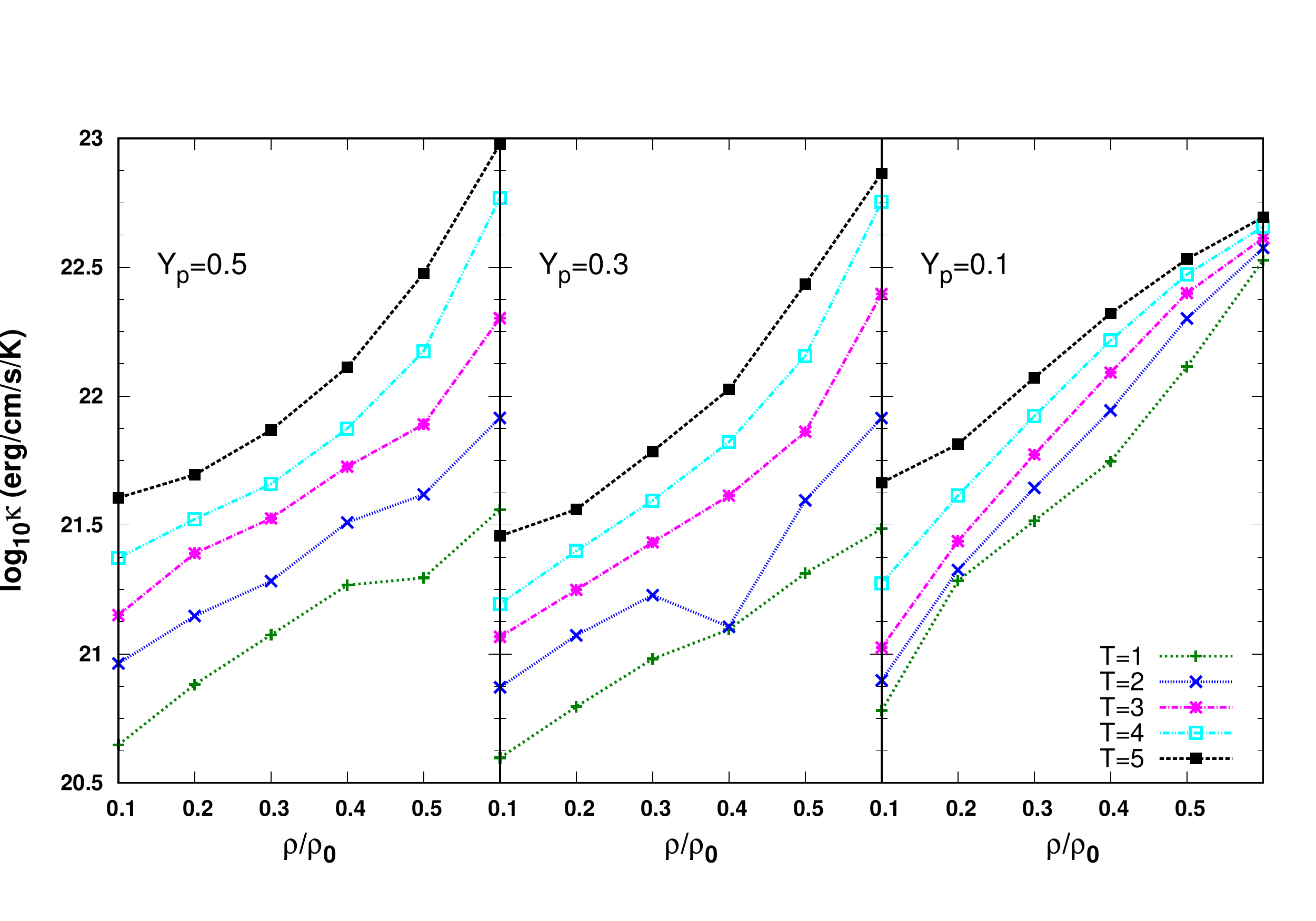}
 \caption{Thermal conductivity as a function of density at different temperatures and proton fractions.}
 \label{fig:kappa}
\end{figure}
\begin{figure}
  \includegraphics[width=0.65\textwidth, angle=-90]{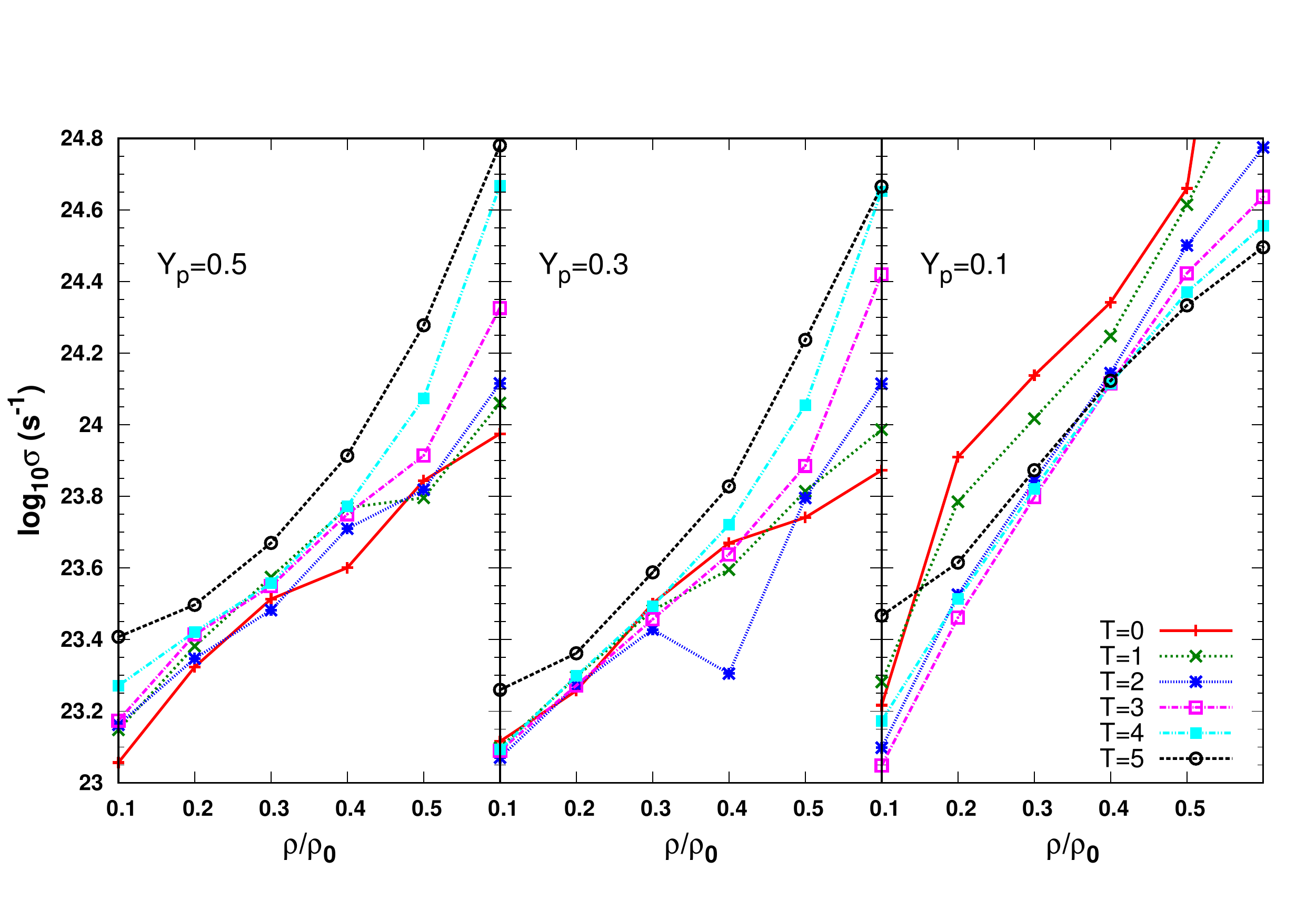}
 \caption{Electrical conductivity as a function of density at different temperatures and proton fractions.}
 \label{fig:sigma}
\end{figure}

In Fig. \ref{fig:eta}-\ref{fig:sigma} we plot the shear viscosity, thermal conductivity and electrical conductivity, respectively as a function of density for
different temperatures and proton fractions. At high densities and/or temperatures the shear viscosity increases smoothly with density as
matter is more or less uniform at these conditions. However, for intermediate densities and temperatures, where the pasta phases appear, the increase in
$\eta$ is not that smooth. In the case of $Y_p=0.3$, we have already noticed in the last section that the occurrence of perfect slabs at $\rho=0.4\rho_0$ and $T=2$ MeV
results in very high values of $S_p(q_{\rm peak})$ and both the $\Lambda_{ep}$s. The plot of $\eta$ also bears this signature as it suddenly decreases at this point.
Except in the transition region of pasta to uniform matter the shear viscosity decreases with temperature at $Y_p=0.1$. This behavior is opposite to the cases of
$Y_p=0.5$ and $Y_p=0.3$ but similar to the results of \citet{Chugonov05}, where the shear viscosity was calculated for the inner crust of neutron stars without 
considering the pasta phases. The values of $\eta$ obtained here have the same orders of magnitude as in \citet{Chugonov05}, suggesting that the presence of
pasta phase does not greatly affect the shear viscosity.

From the Fig. \ref{fig:kappa}, we see that the thermal conductivity increases rather smoothly with density and temperature. Only
at the point of the slab phase there is a dip in $\kappa$ at $Y_p=0.3$. The behavior of the electrical conductivity with density and temperature
is similar to that of the shear viscosity as can be seen from Fig. \ref{fig:sigma}. 
We compare our results for the conductivities of inner crust matter of neutron stars with that of earlier works \citep{Horowitz08, Flowers76, Nandkumar83} .
\citet{Flowers76} presented results for all three transport coefficients of the liquid regime of neutron star matter (which is applicable in our case) 
up to $\sim 10^{11}$ g cm$^{-3}$. When extrapolated to the densities relevant here ($\gtrsim 10^{13}$ g cm$^{-3}$) one gets values similar to us. 
The extrapolation of the results obtained by \citet{Nandkumar83} also gives conductivities of similar orders of magnitude as found in our calculation.

\section{Astrophysical consequences}
As found in the previous section the presence of a pasta phase does not considerably affect the transport properties like shear viscosity and thermal and 
electrical conductivities. This finding has several astrophysical consequences. For example, in a study \citet{Horowitz15} performed
large MD simulation and estimated an impurity parameter ($Q_{\rm imp}\approx40$) for the pasta region in a very simplified fashion. This relatively high value 
of $Q_{\rm imp}$ (without pasta $Q_{\rm imp}\lesssim1$) indicates low thermal conductivity that eventually was used to explain
the late time cooling observed in MXB 1659-29. In another study, \citet{Pons13} considered few values of $Q_{\rm imp}$) for the pasta phase and calculated the 
corresponding electrical conductivities. For $Q_{\rm imp}=100$, the magnetic field decays very fast after $\sim 30000$ years and thereby helps to explain 
the non-existence of isolated X-ray pulsars with spin periods longer than 12 s. 
 
However, the impurity parameter formalism is not actually applicable to the pasta. It was introduced by \citet{Flowers76} to describe a uniform crystal lattice with 
a small fraction of sites occupied by impurities. But, the complicated pasta phase  as found here
cannot be described as a uniform crystal lattice. Moreover, in a study of the outer crust of accreting neutron stars \citep{Daligault09}
it was shown that it is  more accurate to calculate the transport properties directly using  the structure factors than $Q_{\rm imp}$, as the former already
captures all the information of particle correlations. If we do that, the late time cooling of  MXB 1659-29 cannot be explained by the presence of a
pasta phase with our high values of thermal conductivities. Similarly, the high value of the electrical conductivity would fail to explain the absence of X-ray 
pulsars with periods larger than 12 s, which requires a different explanation.  In this context, we plan to perform longer and larger simulations to obtain further quantitative confirmation of these points in future investigations.
We also plan to calculate the neutrino transport coefficients using the obtained structure factors for neutrons ($S_n(q)$) in a upcoming work. 
  
\section{Summary and conclusions}\label{sec:summary}
We have studied the transport properties of nuclear pasta phase within a quantum molecular dynamics approach.
We have performed simulations for a wide range of density ($\rho/\rho_0=0.1-0.6$) and temperature ($T=0-5$ MeV) for this purpose. We have studied
both symmetric nuclear matter, relevant for heavy-ion physics as well as asymmetric matter with $Y_p=0.3$ and
$Y_p=0.1$, important for supernova and neutron star crust environments, respectively. In this context we have computed the thermal and electrical 
conductivities as well as the shear viscosity for all these densities, temperatures and proton fractions. In these conditions electrons are the most 
important carriers of charge and momentum and all the transport coefficients are determined by calculating the Coulomb logarithms that describe
electron-proton scattering. The most important quantity in evaluating the Coulomb logarithms is the static structure factor $S_p(q)$ which describes
correlations between protons. The static structure factors are calculated directly from the particle trajectories obtained in the simulations. The $S_p(q)$ shows a
peak at specific values of $q$, the locations of which is given by the average distance between the nuclear clusters. The peak height $S_p(q_{\rm peak})$ is proportional
to the number of nucleons in the cluster but limited by both the nuclear form factor and screening effects of ions. It is found that in the density and temperature
range of the pasta phase $S_p(q)$ shows irregular behavior. At a few instances we found a sharp rise in $S_p(q)$ due to the presence of almost perfect equidistant slabs.
We also calculate static structure factors for neutrons $S_n(q)$, which we shall use to calculate neutrino transport in core-collapse supernova in a future work.
For the Coulomb logarithms, from which the calculation of transport coefficients is straightforward, we provide fit functions that reproduce the data reasonably
well, which can be implemented in numerical studies like supernova simulations. Although the irregularities in $S_p(q)$ somewhat affects the transport coefficients, the effect is not very dramatic. The shear
viscosity generally increases with temperature at $Y_p=0.5$ and $Y_p=0.3$, but at $Y_p=0.1$ the behavior is the opposite. The electrical conductivity shows
similar features. However, the thermal conductivity increases with temperature at all proton fractions. 
The values of all three transport coefficients are found to have the same orders of magnitude as 
found in theoretical calculations for the inner crust matter of neutron stars without the pasta phase and therefore, contradicts earlier speculations that a pasta 
layer might have low thermal as well as electrical conductivities. We also discuss possible astrophysical consequences of this finding.

\section*{Acknowledgements}
R. N. and S. S. acknowledge financial support from the Helmholtz International Center for FAIR (HIC for FAIR). Major parts of 
the calculations have been performed at the computing facilities of the Center for Scientific Computing at Frankfurt University.

\appendix
\section*{Appendix}
We fit both the Coulomb logarithms $\Lambda_\eta$ and $\Lambda_\kappa$ as a function of $\rho/\rho_0$ for a fixed $T$ using the
functions given in Eqs. (\ref{eq:fit_yp5}) and (\ref{eq:fit_yp1}). All the fit paremeters are presented in Table \ref{tab:fit}.

\begin{sidewaystable}
\begin{center}
\caption{Fit parameters for $\Lambda_\eta$ and $\Lambda_\kappa$ (See Eqs. (\ref{eq:fit_yp5} and \ref{eq:fit_yp1})).}
\label{tab:fit}
{\scriptsize\begin{tabular}{ccccccccccccc}
  \hline
  parameters & \multicolumn{2}{c}{$T=0$} &\multicolumn{2}{c}{$T=1$} &\multicolumn{2}{c}{$T=2$} & \multicolumn{2}{c}{$T=3$} & \multicolumn{2}{c}{$T=4$} &\multicolumn{2}{c}{$T=5$}\\
             &$\Lambda_\eta$&$\Lambda_\kappa$&$\Lambda_\eta$&$\Lambda_\kappa$&$\Lambda_\eta$&$\Lambda_\kappa$&$\Lambda_\eta$&$\Lambda_\kappa$&$\Lambda_\eta$&$\Lambda_\kappa$&$\Lambda_\eta$&$\Lambda_\kappa$\\
  \hline
  \multicolumn{13}{c}{$Y_p=0.5$}\\
  \hline
  $a_0$ &56.8004&63.4073&18.4375&23.0698&7.85166&11.1091&25.8613&29.5707&5.91575&8.09474&-7.3609&-6.49551\\
  $a_1$ &161.635&182.108&0.625201&13.2622&-36.7904&-29.3721&44.4827&54.7427&-24.9493&-19.9177&-70.459&-69.8671\\
  $a_2$ &205.483&230.593&-40.1246&-26.4908&-78.3583&-71.4046&39.7259&51.2689&-44.2851&-39.2154&-107.461&-108.127\\
  $a_3$ &116.262&129.769&-38.2071&-31.6985&-52.4401&-49.4689&17.6567&23.479&-24.034&-21.6038&-60.8372&-61.5885\\
  $a_4$ &23.7091&26.3442&-10.0035&-8.86128&-11.5896&-11.1136&2.93852&4.01932&-4.39697&-3.95405&-11.8864&-12.0747\\
  \hline
  \multicolumn{13}{c}{$Y_p=0.3$}\\
  \hline
  $a_0$ &-3.1533&0.00513832&24.7538&29.9592&85.7989&93.8125&14.2877&18.9362&18.5812&22.5758&-6.60114&-5.13649\\
  $a_1$ &-85.0898&-79.9437&43.6675&58.4075&333.752&361.834&-6.00405&6.6657&11.3364&21.9772&-76.5437&-74.6909\\
  $a_2$ &-156.089&-153.637&44.8381&62.2563&513.657&552.216&-29.9672&-15.8772&-7.27079&4.32513&-122.889&-122.572\\
  $a_3$ &-102.61&-102.743&23.8361&33.1044&324.108&346.704&-21.0457&-13.9772&-7.51401&-1.76271&-71.9138&-72.3447\\
  $a_4$ &-22.5379&-22.7643&4.82296&6.63115&70.3301&75.0142&-4.62655&-3.31657&-1.65026&-0.589162&-14.3787&-14.5317\\
  \hline
  \multicolumn{13}{c}{$Y_p=0.1$}\\
  \hline
  $a_0$ &-0.8779&-0.835066&0.829846&1.0883&1.2638&1.44654&1.41124&1.52231&1.21835&1.29772&0.75379&0.767622\\
  $a_1$ &-0.690231&0.0668335&5.1939&6.43341&5.44127&6.14928&4.86963&5.1891&3.54298&3.69067&1.24756&1.09079\\
  $a_2$ &3.86433&5.78927&10.7386&12.7727&8.85522&9.8851&6.79984&7.23389&4.82357&5.01594&1.73474&1.49322\\
  $a_3$ &4.13683&5.45014&6.85579&7.92397&3.95218&4.33985&1.97647&2.04162&0.845797&0.785266&-0.731935&-1.03983\\
  $a_4$ &1.25123&1.5713&1.53041&1.74534&0.736756&0.815957&0.255757&0.277993&-0.0431151&-0.0601129&-0.440889&-0.524348\\
  \hline
 \end{tabular}}
\end{center}
\end{sidewaystable}

\end{document}